# PTCDA molecular monolayer on Pb thin films: An unusual $\pi$-electron Kondo system and its interplay with a quantum-confined superconductor

Shuangzan Lu[1], Hyoungdo Nam[2], Penghao Xiao[3#], Mengke Liu[2], Yanping Guo[1], Yusong Bai[1], Zhengbo Cheng[1], Jinghao Deng[1], Yanxing Li[2], Haitao Zhou[4], Graeme Henkelman[3], Gregory A. Fiete[5,6], Hong-Jun Gao[4], Allan H. MacDonald[2], Chendong Zhang[1*] and Chih-Kang Shih[2*]

[1]*School of Physics and Technology, Wuhan University, Wuhan 430072, China*

[2]*Department of Physics, the University of Texas at Austin, Austin, TX 78712, USA*

[3]*Department of Chemistry, the University of Texas at Austin, Austin, TX 78712, USA*

[4]*Institute of Physics, Chinese Academy of Sciences, Beijing 100190, China*

[5]*Department of Physics, Northeastern University, Boston, MA 02115, USA*

[6]*Department of Physics, Massachusetts Institute of Technology, Cambridge, MA 02139, USA*

[#]*Present address: Materials Science Division, Lawrence Livermore National Laboratory, Livermore, California 94550, USA*

**cdzhang@whu.edu.cn (C.D.Z)* **shih@physics.utexas.edu (C.K.S.)*

**The hybridization of magnetism and superconductivity has been an intriguing playground for correlated electron systems, hosting various novel physical phenomena. Usually, localized *d*- or *f*-electrons are central to magnetism. In this study, by placing a PTCDA (3,4,9,10-perylene tetracarboxylic dianhydride) molecular monolayer on ultra-thin Pb films, we built a hybrid magnetism/superconductivity (M/SC) system consisting of only *sp* electronic levels. The magnetic moments reside in the unpaired molecular orbital originating from interfacial charge-transfers. We reported distinctive tunneling spectroscopic features of such a Kondo screened π-electron impurity lattice on a superconductor in the regime of $T_K >> \Delta$, suggesting the formation of a two-dimensional bound states band. Moreover, moiré superlattices with tunable twist angle and the quantum confinement in the ultra-thin Pb films provide easy and flexible implementations to tune the interplay between the Kondo physics and the superconductivity, which are rarely present in M/SC hybrid systems.**

The combination of magnetism and superconductivity, which are normally mutually exclusive, provides an intriguing platform involving rich quantum phenomena, such as the Yu-Shiba-Rusinov (YSR) bound states [1-3], the topological superconductivity harboring exotic Majorana modes [4,5], and the heavy-fermion behavior [6,7]. Among the hybrid systems experimentally explored thus far, magnetism is mostly derived from unpaired *d*- or *f*- electrons in transition metal atoms. Compared with *d*/*f* electrons, the *s/p*-electrons show distinctively different behaviors, such as more delocalized wavefunctions (therefore, larger spin correlation lengths) [8] and hyperfine spin-orbit couplings [9,10]. The creation of magnetic properties by $\pi$-electrons has attracted significant interest, which is expected to exhibit improved performance in spin-based information processing [10,11]. Various attempts have been made to achieve magnetism in graphene nanostructures by introducing sublattice imbalance [12-15] or topological frustration [16,17]. Also, charge transfer is another approach to introduce unpaired $\pi$-electrons in pure-organic molecules [18-20]; however, it has strict requirements on the work-function matching. Thus, only scarce examples were reported so far. By either of these two approaches, $\pi$-electron magnetic moments associated with a superconductor have not yet been realized. Moreover, few strategies were known to readily tune the $\pi$-electron magnetism, limiting the in-depth explorations to novel physics within variable regimes of the moment concentrations and the interaction strengths.

Here, we report a hybrid bilayer system comprised of a monolayer (ML) of the organic molecule (3,4,9,10-perylenetetracarboxylic-dianhydride, *i.e.*, PTCDA) and a superconducting Pb thin film. Although none of these two materials contains magnetism, surprisingly, we found that net spin moments formed in the molecular film, resulting in Kondo resonances near the Fermi level. First-principle calculations support the formation of spin-polarized lowest unoccupied molecular orbital (LUMO) states induced by interlayer charge transfer. By scanning tunneling spectroscopy investigations, we revealed distinctive characteristics of the combination of a Kondo screened $\pi$-orbital impurity lattice and a superconductor. In particular, our studies suggest the formation of a two-dimensional (2D) Kondo-induced impurity band near the superconducting gap edge. More interestingly, we found twistable moiré superlattices forming in this bilayer system, leading to moiré modulations for the Kondo-superconductivity interplay. In addition, the quantum confinement effect in Pb films provides another tuning knob to the charge transfer induced magnetic moments. These two appealing features combine in a single sample system, manifesting versatile tailoring of the complex interactions at the M/SC interface.

Figure 1(a) is a scanning tunneling microscopy (STM) image showing a crystalline PTCDA layer formed on a Pb(111) film (23 ML here), which is grown epitaxially on Si(111) (Methods in [21]). The herringbone structure of a PTCDA layer has a rectangular unit cell with unit vectors $\vec{\boldsymbol{a}}_1$= 1.29 ± 0.02 nm and $\vec{\boldsymbol{a}}_2$= 1.81 ± 0.02 nm. The angle between $\vec{\boldsymbol{a}}_1$ and Pb<$1\bar{1}0$> is defined as $\theta$ in the inset of Fig. 1(a). In Fig. 1(b), we show a typical tunneling spectrum (blue curve) taken on the ML PTCDA/Pb(111) film at a sample temperature $T_S$ above the $T_C$ of Pb film. A resonance peak appears at the Fermi level. The superconducting gap can coexist with this resonance peak (grey curve) when $T_S < T_C$. Based on the temperature dependence of the resonance [21] and its interplay with the superconductivity, we believe this resonance resulted from the Kondo effect. Note that the monolayer PTCDA with the herringbone structure can form on many metallic surfaces, yet no study has reported the observation of Kondo resonances in the pristine molecular film [20,45]. To support this hypothesis, we grew a PTCDA monolayer on the Ag(111) surface and found that the Kondo resonance was absent [red curve in Fig. 1(b)]. In addition, the Kondo resonance was absent for a single molecule on the Pb film [green curve in Fig. 1(b)] but emerged only after in-plane molecular hybridization occurred.

To understand the origin of the local magnetic moment, we performed DFT calculations for a single PTCDA molecule/Pb(111), ML PTCDA/Pb(111), and ML PTCDA/Ag(111) [21]. In Fig. 1(c), we plot the charge transfer and the magnetic moment for each molecule as a function of the interlayer separation *d* for all three systems. The calculations show a result consistent with the experimental observation that only monolayer PTCDA on Pb(111) possesses magnetic moments at the equilibrium distance (marked by vertical dashed lines). Figure 1(d) shows the partial density of states (DOS) of ML PTCDA/Pb at four interlayer separations *d*. As shown, the intra-orbital Coulomb repulsion energy *U*, the impurity state (*i.e.*, singly occupied LUMO) bandwidth *W*, and energy level ($\varepsilon_{imp}$) all varied with *d*. Note that the LUMO states comprise relatively delocalized *s* and *p* electrons, leading to a small magnitude of $U$ ~ 0.1 eV, which is comparable to the $W$ and $\varepsilon_{imp}$. This explains why the DFT magnetic moment does not always scale linearly with the charge transfer in Fig. 1(d). Detailed discussions are given in the SM [21].

The quantitative strength of the Kondo resonance is strongly modulated by the moiré superlattices forming between the molecular and the Pb lattices. Such moiré structures show long-range ordered super-periodicities and a tunable twist angle $\theta$, both of which are absent in previous observations of the adsorption site dependent Kondo effect [27,36]. Figure 2(a–b) show three typical moiré superstructures with $\theta = 11°$, $\lambda = 10a_1$; $\theta = 13°$, $\lambda = 8a_1$; and $\theta = 17°$, $\lambda = 6a_1$ where $\lambda$ represents the moiré periodicity along the $\vec{a}_1$ direction. The upper panel in Fig. 2(c) shows a color rendering of the d$I$/d$V$ spectra for 20 molecules along a bright row as labeled in Fig. 2(a). These d$I$/d$V$ spectra can be fitted with the widely adopted Fano formula [25]. Representative spectra and the corresponding fits are displayed in Fig. S3(a). In the lower panel of Fig. 2(c), we show a plot of the half-width at half-maximum of the resonance ($\Gamma$) for each molecule. The peak value occurs at the location with the lowest topographic height (index #10 and #20), while the minimum value occurs at the highest locations (index #6 and #15). In all our measurements for the moiré superlattice with $\theta = 11°$, $\Gamma$ varies from a minimum of 6.3 meV to a maximum of 20.5 meV, corresponding to a variation of $T_K$ from 72 K to 238 K. A similar spatial mapping of the Kondo resonance for $\theta = 13°$ is displayed in Fig. S5. Both the oscillation amplitude ($\Gamma$ =

8.8 meV - 14.9 meV) and the average magnitude of Kondo screening are significantly smaller than that for $\theta = 11°$, demonstrating a twist-angle tuned Kondo screening for 2D impurity lattice. In this work, all spectra are acquired at the C-H bond location [21] when discussing the inter-molecular difference, unless stated otherwise.

When a single localized magnetic moment is coupled to a superconductor, its exchange interaction with the Cooper pairs leads to in-gap states, referred to as YSR bound states, after the seminal works of Yu, Shiba, and Rusinov [1-3]. Matsuura extended the original framework to include the Kondo screening and identified two regimes: (a) “free spin” regime ($\Gamma < \Delta$) where Kondo screening is quenched; and (b) “spin-screen” regime ($\Gamma > \Delta$) where the efficient Kondo screening leads to a singlet ground state. In the "spin screen" regime, the quasiparticle excitations are strictly not YSR states, albeit this distinction has been blurred in several recent literatures [46,47]. To avoid confusion, we simply refer to the quasiparticle excitations in the $\Gamma \gg \Delta$ regime as “bound states” [40,48]. Exchange interactions of local moments and superconductors have gained tremendous interest lately. Nevertheless, most STS investigations have focused mainly on single magnetic impurities or impurity pairs [49-51]. Here we present a regime with the interaction of “Kondo screened delocalized 2D spin lattices” and the 2D superconductors where qualitatively different behaviors are observed.

Figure 3(a) shows d$I$/d$V$ spectra (−4.0 mV to +4.0 mV) for three representative molecules [index #5, #6, and #10 in Fig. 2(c)] taken with an SC Nb tip at 0.4 K. These spectra exhibit two sharp peaks (hole-particle symmetry in energy) each accompanied by a dip right next to them. By contrast, the spectrum acquired on the Pb surface does not exhibit such dips. After deconvolving of the Nb tip DOS [21], the sample DOS are displayed in Fig. 3(c). At first sight, these sample DOS resemble the pristine superconducting gap, except with smaller gap values. However, numerical analysis reveals that the presence of dips in Fig. 3(a) is related to spectral weight conservation, which manifests the formation of bound states [21]. In spectrum #10, the area of the apparent peak above the normalized reference line is marked as $S_1$ with the "absence" of spectral weight below the reference line denoted as $S_2$. The dip size is directly proportional to the ratio $S_1/S_2$. For a Bardeen-Cooper-Schrieffer (BCS) superconductor, spectral weight conservation ensures $S_1/S_2 = 1$; thus, no such dip would exist. The $S_1/S_2$ ratio represents the extent that the DOS deviates from the BCS line shape. The particle/hole quasiparticle excitation peaks are labeled as $\pm\Delta^*$, with $\Delta^*$ corresponding to the binding energy of the bound state.

A few important characteristics are noted: (i) the absence of multiplets typically observed for a high spin magnetic moment; (ii) a nearly symmetric particle/hole spectral weight; (iii) a spectral width broader than the typical bound state or YSR states; and (iv) a moiré modulation of $\Delta^*$ [Fig. 3(c)] which is in phase with the spatial moiré modulation of $\Gamma$ [Fig. 2(c)]. Interestingly $\Gamma$ *vs.* $\Delta^*$ (including more than 150 molecules over the surface) can be well-fitted with the theoretical model of Matsuura for $T_K >> \Delta$ [40]:

$$\Delta^* = \Delta_0 \frac{1-\alpha^2}{1+\alpha^2}, \text{ with } \alpha = \frac{\Delta_0}{\Gamma} \ln(\frac{\Gamma}{\Delta_0} \cdot e),$$

using an asymptotic value of $\Delta_0^{PTCDA/23ML}$ = 1.06 meV at $\Gamma = \infty$ [Fig. 3(d)]. Here $\Delta_0$ is smaller than the gap value for 23ML Pb film (1.31 meV), suggesting that this hybrid 2D system has a reduced gap. The reduction in the SC gap is confirmed by the proximity effect [Fig. S9(c)], where the bare Pb surface experiences a gradual gap reduction as one laterally approaches a PTCDA island. We attribute this diminution of the SC order parameter to the finite magnetic impurity concentration, as proposed in the original model by Matsuura [48], which is beyond the YSR picture in the dilute limit.

The absence of multiplets in spectra acquired using a superconducting tip [Fig. 3(a)] simply reflects that our system is a 2D lattice comprised of quantum spins. The lack of particle/hole spectral weight asymmetry deserves some special attention. In most previous studies on bound states, the spectral weight between hole-like and electron-like quasiparticle excitations is highly asymmetric. In all our measurements, the present system comprised of $\pi$-electron magnetic moments, the particle/hole spectral weights are nearly symmetric, with a ratio of 0.93 in the most asymmetric case. Two controlling parameters influence the particle/hole quasiparticle spectral function [52]: the spin-exchange interaction ($J$) and Coulomb scattering potential ($K_U$), which contributes to the pair breaking and the particle-hole symmetry breaking, respectively [53]. Adopting the theory in Ref. [39, 54], we can extract $J$ and $K_U$ by analyzing $\Delta^*$ and the particle/hole spectral weight ratio (details in [21]). The moiré oscillations of $N_F|J|$ and $\frac{K_U}{|J|}$ are displayed in Fig.

2(d). The largest value of $\frac{K_U}{|J|}$ (~0.05) occurs at the location with the smallest gap, while on other sites, the values of $\frac{K_U}{|J|}$ are even smaller. The weak potential scattering is a characteristic feature of the magnetic moments residing in the relatively delocalized $\pi$-orbitals, yielding the nearly symmetric particle/hole spectral weights.

The delocalized nature of the spin-polarized orbital is also responsible for the broad spectral width (~0.5 meV in [21]) for the bound states. Further analysis is presented in Fig. S8, where the d$I$/d$V$ spectra mapping with a fine step size shows that the bound states and Kondo resonance are continuous throughout the whole 2D interface, existing even on the sites between molecules. This is fundamentally different from the case of the MnPc/Pb system, where the bound states occur only on the Mn atoms [36]. Note that in the fitting of the Nb tip -Pb tunneling spectrum, small Dynes broadening parameters are used, and the effective temperature is determined as 0.63 K, which is close to the temperature reading (*i.e.*, 0.4 K) [21]. Thus, the large width of bound states can not be attributed to the instrumental energy resolution. As predicted in early theoretical works [3,55], an “impurity band” with a finite bandwidth of the bound states can form when the locally excited states can overlap with each other. Our observation suggests the formation of a 2D bound state band over the Kondo screened impurity lattice. The lateral coupling of the Kondo-screened bound states does not change the basic picture of the single-impurity Kondo effect until the moment concentration is of the same order of magnitude as the free carrier concentration [56]. We hope our discovery will inspire future theoretical efforts on this mixed regime.

Finally, we show that the quantum confinement effect (QCE) in Pb films can dramatically affect the Kondo/superconductivity interplay. Figure 4(a) shows large-energy-scale d*I*/d*V* spectra taken on 22ML and 23ML Pb films for both the bare Pb and PTCDA/Pb hybrid system. The thickness-dependent quantum well states are clearly seen [22,41,57,58]. The Γ values were 17.4 meV and 38.1 meV for the highest and lowest molecules within a moiré periodicity on 22ML Pb [Fig. 4(b)], which represents an enhancement by a factor of 2 in the Kondo screening energy and moiré modulation amplitude, compared with the results on 23 ML. The corresponding values of $\Delta^*$ were 0.97 meV and 1.16 meV, respectively [Fig. 4(c)]. The variation of $\Delta^*$ as a function of Γ on PTCDA/22ML-Pb yields an asymptotic $\Delta_0^{PTCDA/22ML}$ of 1.23 meV [Fig. 3(d)], which is significantly larger than the value of 1.06 meV for PTCDA/23ML Pb. The QCE only leads to a 3% change in SC transition temperature $T_C$ between 22 ML and 23 ML bare Pb [21, 22]. After the Pb films were covered with a monolayer of PTCDA, the pairing strengths (fully Kondo-screened) show a difference of 15% (1.23 meV *vs.* 1.06 meV). This observation illustrates the complex and intriguing interfacial interplay in our system, where the thickness-dependent work function (and the DOS at $E_F$) [41] dictates the average magnetic moment and then determines the “impurity” concentration and the screening strength by the surrounding itinerant electrons [21], which in turn facilitates a much more effective tuning of the SC pairing strength.

In conclusion, we have established an unusual magnetism/superconductivity hybrid bilayer with pure $\pi$-electrons that are carrying the magnetic moments. Owing to the relatively delocalized nature of $\pi$–electrons, this hybrid system presents 2D-like behaviors for the interlayer stacking registry, the Kondo screening, and the bound state formation, which qualitatively distinguishes it from *d*-electron- and *f*-electron-based moment systems and thus opens new avenues for novel correlated physics. Moreover, the control over the twist angle offers a rare opportunity to explore the physics of the moiré modulated Kondo and superconductivity. In addition to combining moiré physics with quantum confinement effects, we demonstrated a practical tuning approach for the magnetic moment concentration and the SC pairing strength. Rich emergent quantum phenomena are anticipated with the combination of all these intriguing elements, as well as possibilities for fabrications of 2D organic structure [59] hosting topological and exotic fractionalized states.

We acknowledge funding from the National Science Foundation through the Center for Dynamics and Control of Materials: an NSF MRSEC under Cooperative Agreement No. DMR-1720595 and NSF Grant Nos. DMR-1808751, DMR-1949701, DMR-2114825; the Welch Foundation F-1672 and F-1841. We also acknowledge computing time from the Texas Advanced Computing Center. C.D.Z and H.-J.G thank the supports from the National Key R&D Program of China (Grant No. 2018FYA0305800 and 2018YFA0703700), the National Natural Science Foundation of China (Grant No.

11774268, 11974012 and 61888102), the Ministry of Science and Technology of China (No. 2013CBA01600) and the Strategic Priority Research Program of Chinese Academy of Sciences (Grant No. XDB30000000).

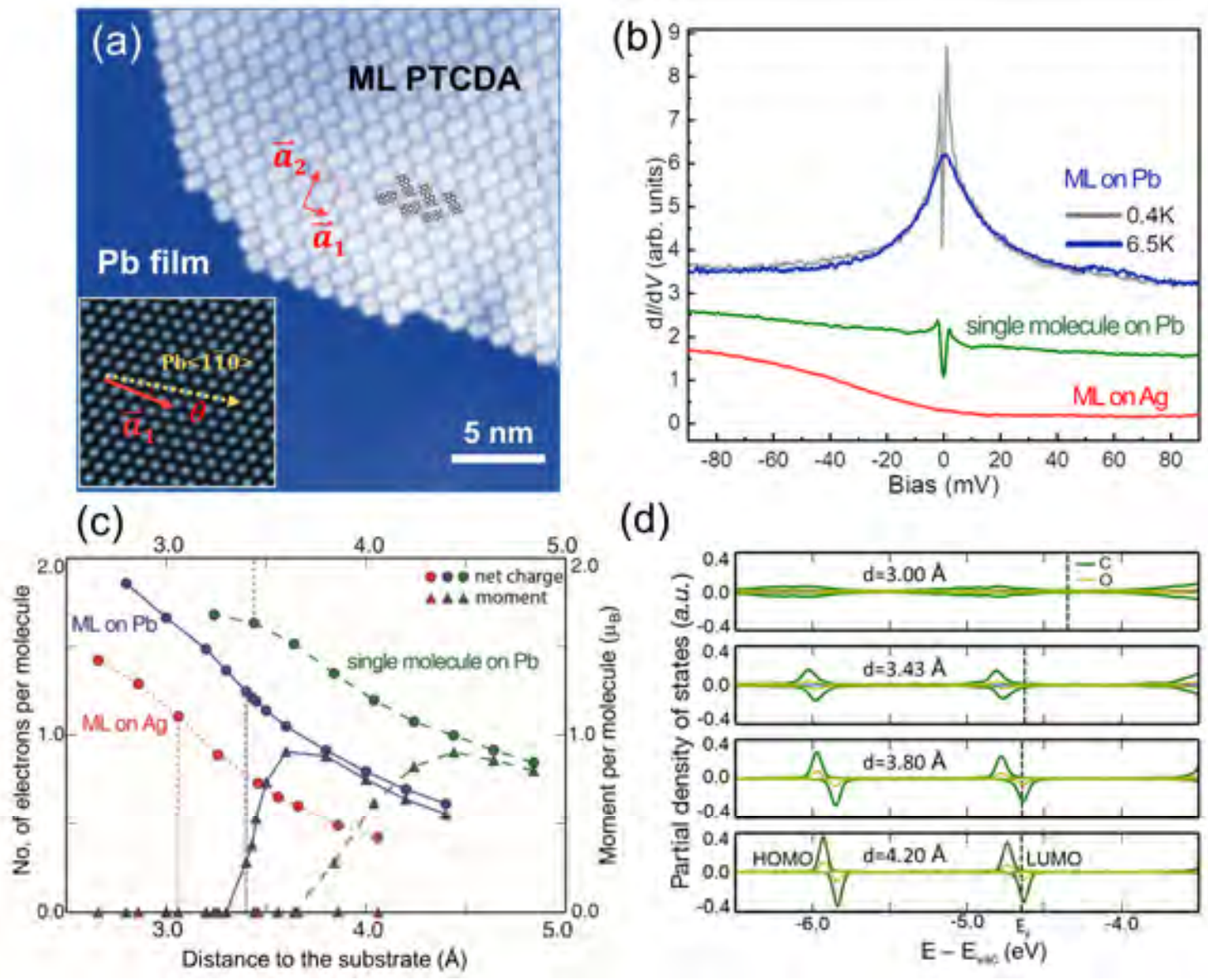

FIG. 1 (a) STM image of ML PTCDA/Pb film. The red rectangle represents a unit cell of the herringbone structure. The inset shows the atomically resolved image of the nearby Pb surface (4×4 nm$^2$). The twist angle $\theta$ is defined as labeled. (b) d$I$/d$V$ spectra for the ML PTCDA/Pb, $T_S$ = 0.4 K (gray) and 6.5 K (blue), a single molecule on Pb at 0.4 K (green), and the ML PTCDA on Ag(111) at 4.2 K (red). Spectra are shifted vertically for clarity. (c) The charge transfer and the magnetic moment as a function of the $d$ for all three situations. The equilibrium distances are marked by vertical dashed lines. (d) Partial DOS at various $d$ for ML PTCDA/Pb. The dashed lines indicate the Fermi levels (a) Sample bias $V_s$ = 1.0 V, set-point current $I$ = 25 pA. Setpoint in (b): $V_S$ = 95 mV, $I$ = 200 pA, and the lock-in modulation $V_{rms}$ = 0.8 mV.

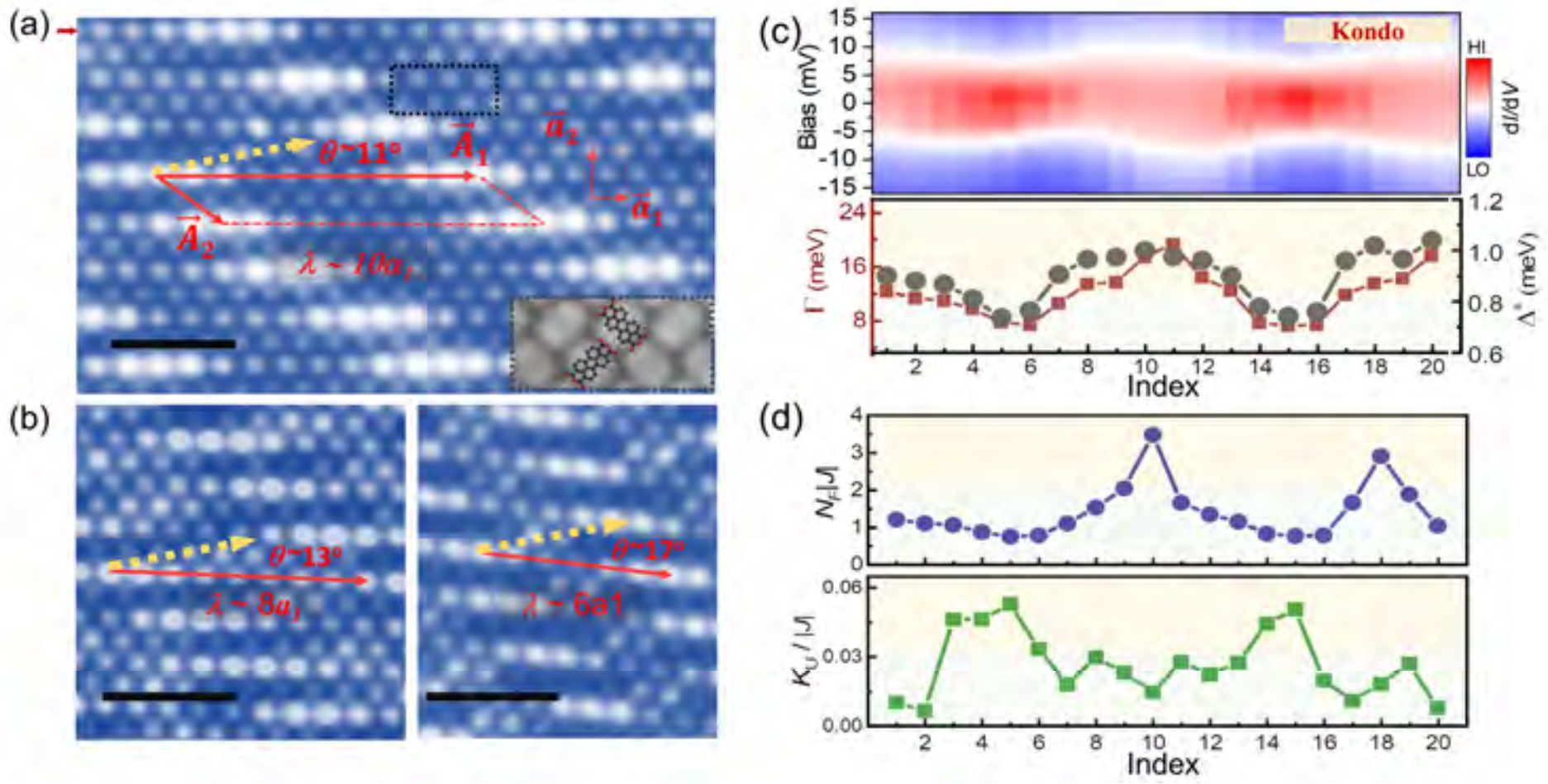

FIG. 2 (a)–(b) Typical Moiré superlattice for $\theta = 11^{\circ}$, $\theta = 13^{\circ}$, and $\theta = 17^{\circ}$. The red parallelogram in (a) represents the moiré supercell. $\lambda$ is the periodicity along $\vec{\boldsymbol{a}}_1$ direction. The Pb<$1\bar{1}0$> direction is marked by the yellow dashed arrow. Inset in (a) is a zoomed-in image at the black dashed rectangle with molecular models overlaid. (c) The upper panel is a false-color image of d$I$/d$V$ spectra taken along the bright row marked in (a) by the red arrow. The lower panel shows the moiré modulations of Γ (wine) and Δ* (gray). A small magnetic field (0.5 T) was applied to quench superconductivity. (d) The corresponding oscillations of $N_F|J|$ (upper panel) and $K_U/|J|$ (lower panel). Spectra in (c) were taken at 0.4 K with $V_s = 95$ mV, $I = 100$ pA, $V_{rms} = 2.5$ mV. (a-b) $V_s = 2.0$ V, $I = 20$ pA; inset in (a) $V_s = 1.0$ V. Scale bars are 5.0 nm.

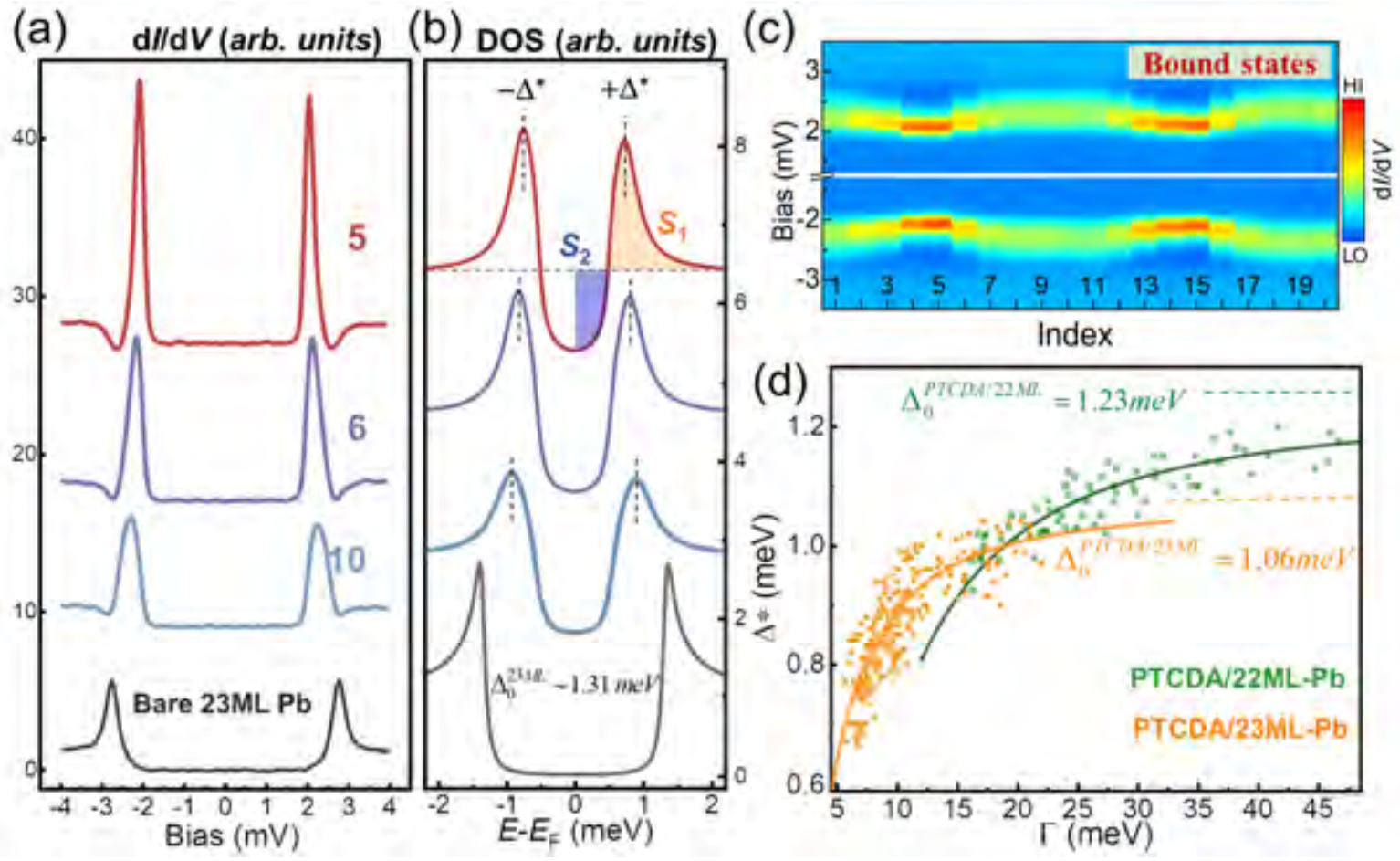

FIG. 3 (a) Pairing gap measurements for the molecule #5, #6 and #10 in the bright row in Fig. 2(a). (b) Sample DOSs obtained by numerical deconvolution of the superconducting tip DOS. Results of bare 23ML Pb are displayed for comparison. The peak area $S_1$, gap area $S_2$ and bound state energy level $\Delta^*$ are labeled as shown. (c) False-color image of the bound state measurements along the bright row in Fig 2(a). (d) Ensembles of data points on PTCDA/23ML Pb (orange) and PTCDA/22ML Pb (green), plotted in terms of the $\Delta^*$ versus $\Gamma$. The solid lines are fittings based on Matsuura's theory. The asymptotic limits of the SC order parameters at $\Gamma = \infty$ are labeled. All spectra were taken by Nb tips at 0.4 K. $V_s = 4.0$ mV, $I = 100$ pA, $V_{rms} = 50\mu$V.

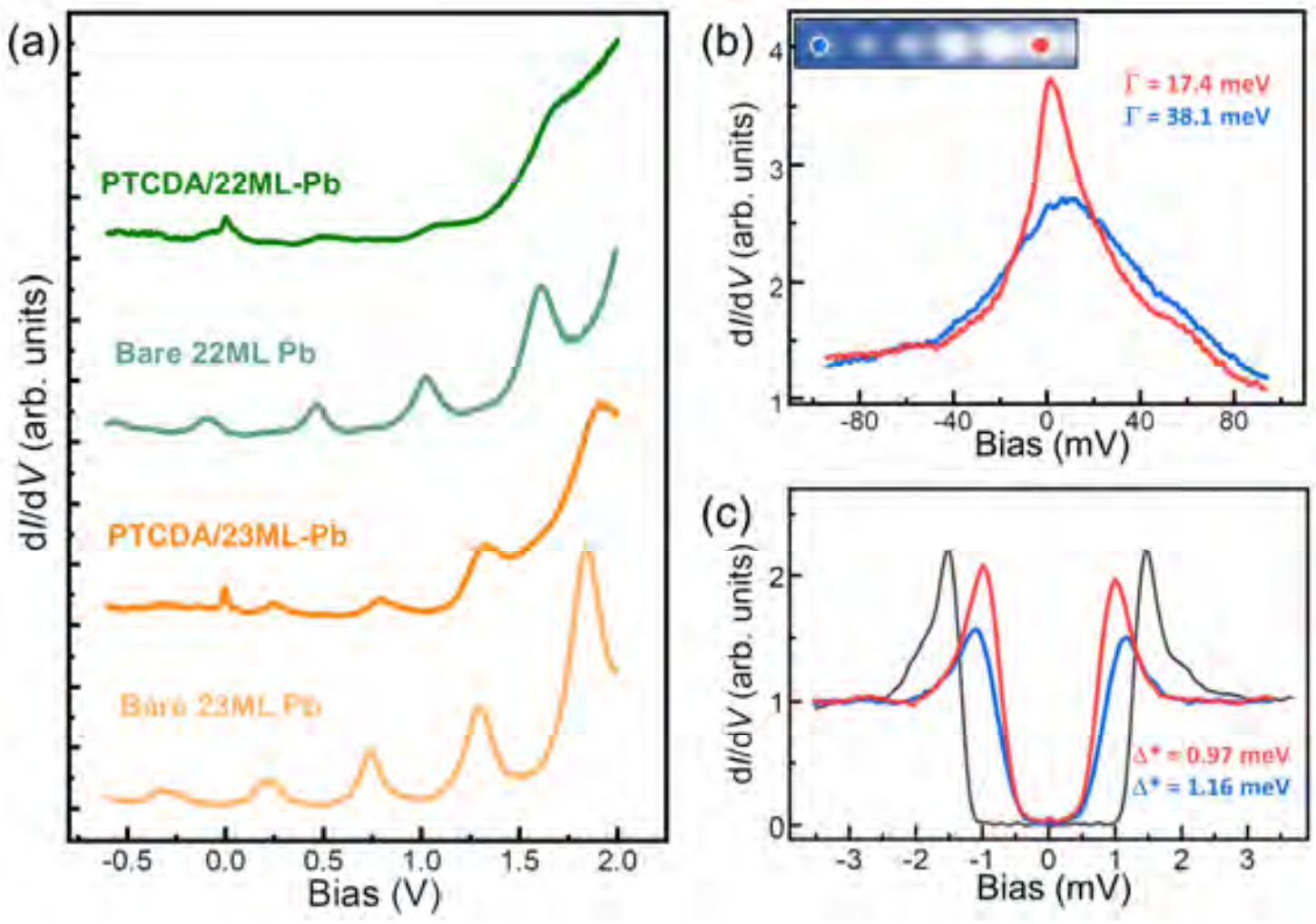

FIG. 4 (a) d$I$/d$V$ spectra from -0.6 V to +2.0 V taken on the ML PTCDA/22 ML Pb, bare 22 ML, ML PTCDA/23 ML Pb, and 23 ML Pb surfaces. (b) and (c) Measurements of Kondo resonances and pairing gaps on ML PTCDA/22 ML Pb. Two typical spots are chosen: the morphologically highest (red) and lowest (blue) molecules within one moiré periodicity [labeled in the inset of (b)]. The black curve in (c) is taken on bare 22 ML Pb with its SC gap $\Delta_0^{22ML}$=1.35 meV [21]. All spectra were taken at 0.4 K with a normal tip. (b) $V_s$ = 95 mV, $I$ = 100 pA, $V_{rms}$ = 2.5 mV; (c) $V_s$ = 3.5 mV, $I$ = 100 pA and $V_{rms}$= 50 μV.

# Supplemental Material for

# PTCDA molecular monolayer on Pb thin films: An unusual $\pi$-electron Kondo system and its interplay with a quantum-confined superconductor

Shuangzan Lu[1], Hyoungdo Nam[2], Penghao Xiao[3#], Mengke Liu[2], Yanping Guo[1], Yusong Bai[1], Zhengbo Cheng[1], Jinghao Deng[1], Yanxing Li[2], Haitao Zhou[4], Graeme Henkelman[3], Gregory A. Fiete[5,6], Hong-Jun Gao[4], Allan H. MacDonald[2], Chendong Zhang[1*] and Chih-Kang Shih[2*]

**List of contents:**

## 1. Materials and Methods

Clean Si(111)-7×7 surface was prepared by cycles of flashing at 1480 K using direct-current heating (heavily n-type doping with a resistivity < 0.005 Ω/cm). A commercial Knudsen effusion cell was used to evaporate Pb (0.35ML/min), while a home-built evaporator was used for the PTCDA molecules (purchased from *Aldrich Inc.*) with a deposition ratio of 0.4 ML/min. Pb films were grown on Si(111) surface directly without preparation of the crystalline wetting layer. It was first deposited at a sample temperature of ~100K [22]. The thickness of Pb film was monitored by a quartz crystal monitor and was examined by energy levels of the quantum wells states. Varied annealing procedures were carried out to prepare different moiré structures. We found that the immediate sample transferring with minimum temperature increasing can lead to the formation of a moiré pattern with $\lambda = 10a_1$. The longer the annealing of the sample, the shorter the $\lambda$ is. The moiré superlattice with $\lambda = 6a_1$ is obtained after fully annealing at room temperature (> 1 hour).

The STS measurements at 0.4 K were conducted by using a Unisoku-1300 STM at Wuhan University. The system can be cooled down to 0.4 K by a single shot $^3$He cryostat; the magnetic field can be applied perpendicularly to the sample surface. Polycrystalline Pt-Ir and Nb STM tips were used in our experiments. The bias voltage is applied to the sample. d*I*/d*V* spectra were measured by using the lock-in technique with the reference signal at 963 Hz. Different modulation amplitudes were used depending on the energy range, as specified in each figure's captions.

## 2. Additional information for the molecular lattice structure

Monolayer PTCDA molecules on Pb(111) film form a well-known herringbone structure with a rectangular unit cell. The unit vectors $\vec{\boldsymbol{a}}_1$ and $\vec{\boldsymbol{a}}_2$ are marked in Fig. S1. When scanning with relatively small sample bias 0.1 V [Fig. S1(a)], each PTCDA molecule is imaged as two elongated maxima (two bright edges). With a sample bias of 1.5 V [Fig. S1(b))], the molecules are imaged as bright protrusions, and the moiré periodicity becomes visible. With special tip conditions (a functionalized tip), we can obtain sub-molecular resolution images [Fig. S1(c)-(e)]. The molecular morphologies taken with +/- 50 mV [Fig.S1(c)-(d)] resemble the previous theoretical and experimental results of the PTCDA LUMO. It worth noting that the PTCDA LUMO states on other surfaces or in the gas phase usually locate above the Fermi level by ~ 1.5 eV [23,24].

However, in our case, it can only be imaged with a small bias and diminishes when the bias is high [0.2 V in Fig. S1(e)]. Such observations proved the picture that the magnetic moments reside in the LUMO electrons whose energy levels are shifted to ~ $E_F$ due to the charge transfer. Thus, the order of U can be estimated as ~ 0.1 eV, which qualitatively agrees with the DFT results. Moreover, with the sub-molecular resolution images, we determine that the bright protrusions of the molecule in Fig. S1(a) correspond to C-H bonds.

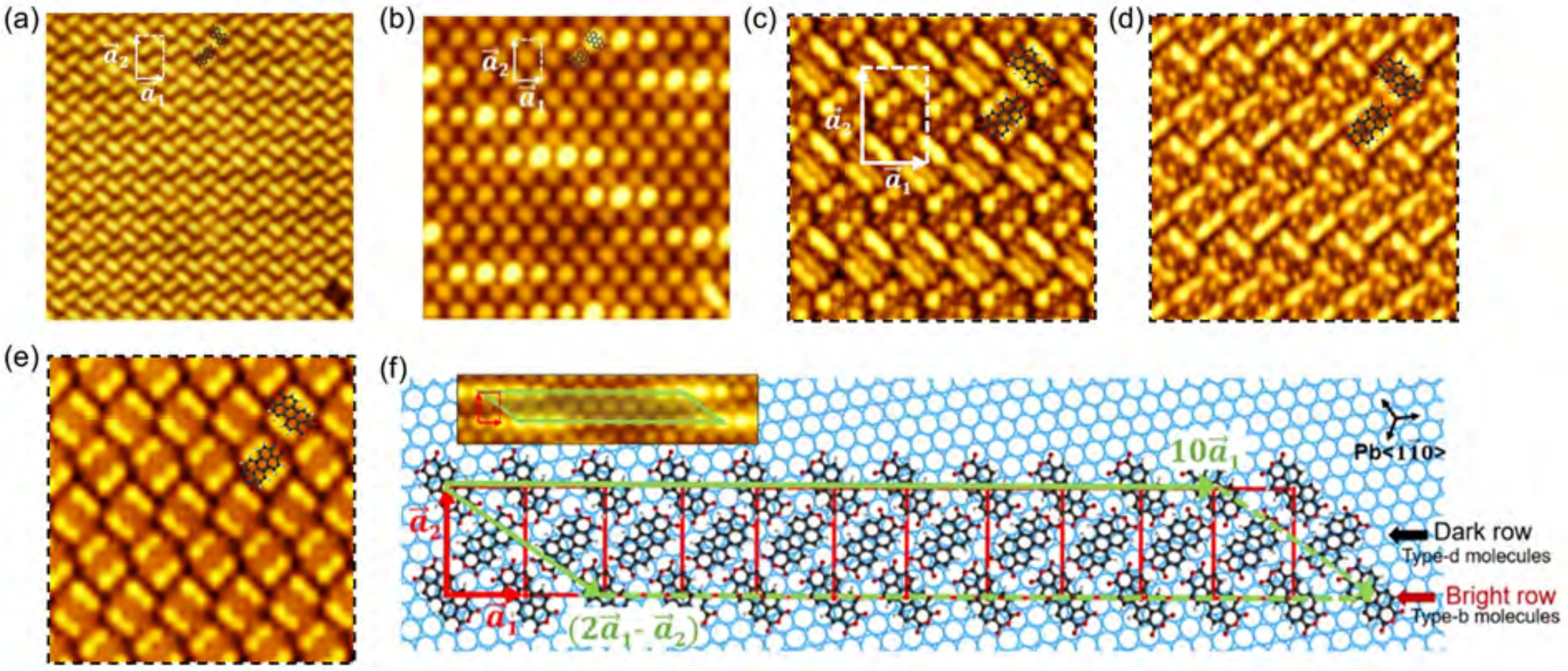

**Fig. S1** (a) and (b) STM images of the PTCDA lattice. (a) $V_s$ = 0.1 V, $I$ = 25 pA; (b) $V_s$ = 1.5 V, $I$ = 50 pA; images size 14 × 14 nm$^2$. (c)-(e) High resolution images taken with a possible functionalized tip. $V_s$ = 50 mV for (c), $V_s$ = -50 mV for (d) and $V_s$ = 200 mV for (e), size 7×7 nm$^2$. Unit cells are labeled by white arrows. Ball-stick models of the molecule are displayed in (a-e). (f) Schematic of a supercell of the moiré pattern (green parallelogram) with $\vec{\boldsymbol{A}}_1 = 10\vec{\boldsymbol{a}}_1$, $\vec{\boldsymbol{A}}_2 = 2\vec{\boldsymbol{a}}_1 - \vec{\boldsymbol{a}}_2$ with a twist angle of $\theta$ = 11°. The Pb(111) surface is shown by closed packed balls in blue. The insert shows a corresponding STM image. The lateral registry between molecular lattice and Pb lattice is not rigorous.

A schematic model of the moiré superlattice at $\theta$ = 11° is shown in Fig. S1(f). The unit-cell of the herringbone structure is marked by a red rectangle containing two molecules (type-b and type-d as labeled). Combined with the atomically resolved image taken on the nearby Pb surface, we determine that the type-b molecules have their long axes nearly aligned with respect to the close package direction (*i.e.*, Pb<1$\bar{1}$0>), while the type-d molecule has the long axis roughly aligned along with Pb<112>. In the moiré superlattice, the type-d molecules form the dark-rows while the type-b molecules form bright-rows [Fig. S1(f)].

As shown in Fig. S2(a), the small molecular clusters (or a single molecule) can be separated from the compact herringbone film by the tip stimuli. The single molecule always adsorbs on the most favorable sites, thus exhibits no Kondo resonance. It is irregular that the registries of the dimer and the few molecules clusters with respect to the Pb lattice along with their inter-molecular configurations. Correspondingly, the occurrences of Kondo resonances on these small irregular clusters are also random [see Fig. S2(b) and color dots in Fig. S2(a)]. Once the ordered herringbone structure is formed, the Kondo will exhibit all over the cluster. So far, the smallest cluster we observed to have the herringbone structure contains four molecules.

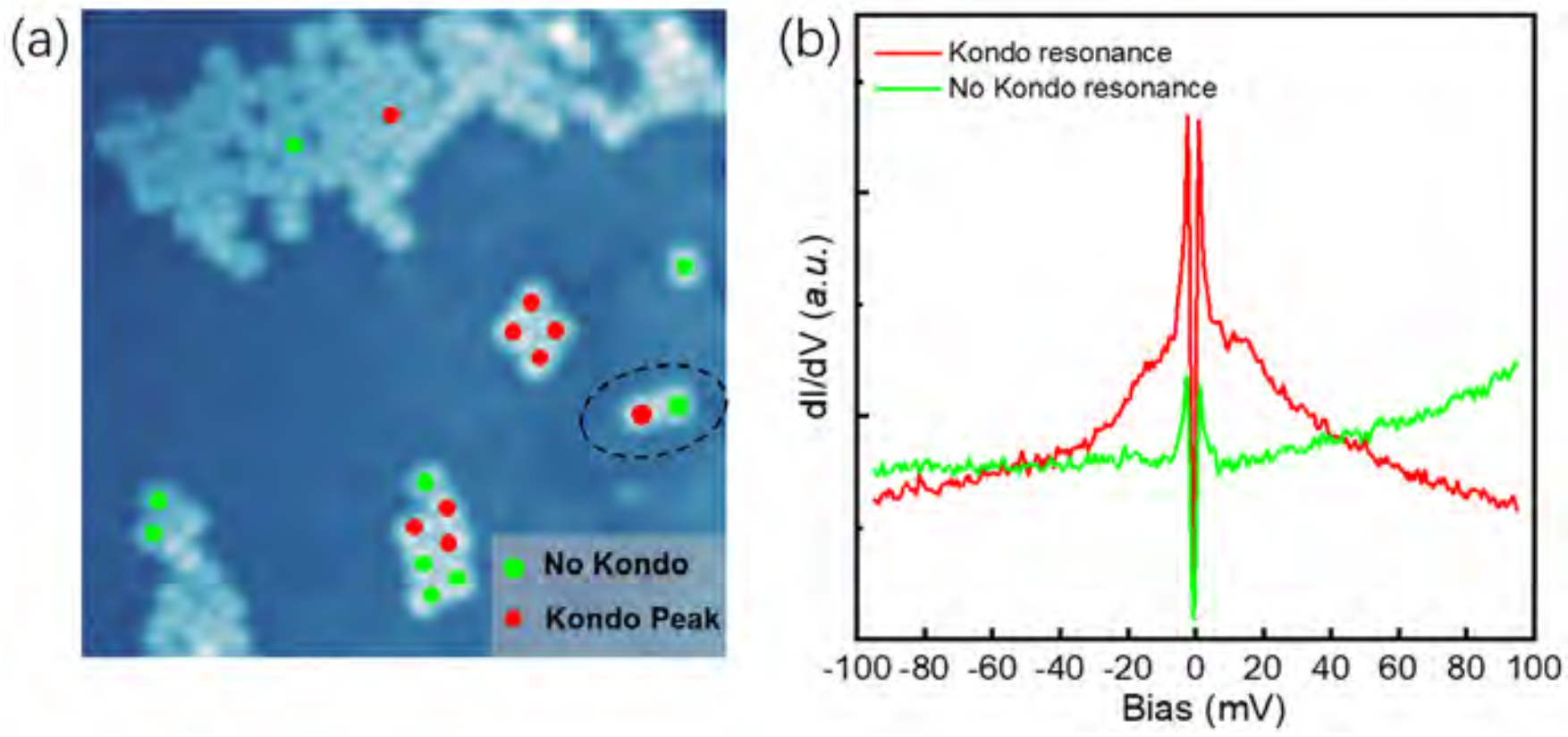

**Fig. S2** (a) Single molecule, few molecules, and irregular small islands separated from a packed film by the STM tip. The existences of Kondo resonance are marked by the red (with Kondo) and the green (w/o Kondo) dots. $V_s$ = 0.2 V, $I$ = 40 pA. (b) Representative spectra obtained on the molecule dimer marked by the black dashed circle in (a). Feedback opened at $V_s$ = 95 mV, $I$ = 300 pA and $V_{rms}$ =1.5 mV.

## 3. Details of the first-principles calculations

(1) **Calculation methods.** The density functional theory calculations were performed under the generalized gradient approximation (GGA) scheme with the Perdew–Burke–Ernzerhof (PBE) functional using the Vienna ab-initio simulation package (VASP). Core electrons were described using the projector augmented wave (PAW) framework. The softer version of the PAW potentials was used for C and O to save computational time. The energy cutoff for the plane-waves basis set was 283 eV. The Brillouin zone was sampled on a 2×4×1 Monkhorst-Pack mesh. The Tkatchenko-Scheffler with self-consistent screening (TS+SCS) method was used to account for vdW interactions. All calculations were spin-polarized. A dipole correction was applied to cancel the electrostatic interaction between the slab surfaces due to charge transfer. The atomic configurations of a PTCDA monolayer on the metal substrate were first fully relaxed by converging the interatomic forces below 0.03 eV/Å. To calculate the dependences on $d$, the molecular monolayer or single molecule was then manually translated without further structure relaxation to obtain the DOS at other distances. The distance to the substrate is reported as an average over all atoms. The vertical dashed lines in Fig. 1(c) indicate the equilibrium molecule-substrate distances, which are 3.46 Å for the single-molecule/Pb, 3.43 Å for the molecules monolayer/Pb, and 3.06 Å for the molecules monolayer/Ag. In all DOS plots, the vacuum energy level is set to zero. The partial densities of states of the ML PTCDA/Pb system as a function of separation distance are plotted in Fig. 1(d). Figure S3(a) shows the spin density mapping of the ML PTCDA/Pb system at their equilibrium distance. The spin-up (yellow) cloud resembles the LUMO state of a gas phase PTCDA. This further confirms that the local moment is due to the charge transferred from Pb to PTCDA. We'd like to clarify that due to the limitations on supercell size in calculations, The DFT results here can be used only as a guide for qualitative trends. The formation of net magnetic moment in the LUMO states is a consequence of the competition between the charge transfer and the hybridizations of molecular orbitals.

(2) **Discussions for the formation of magnetic moments.** As seen in Fig. 1(c), the charge transfer $Q$ always scales linearly with the interlayer distance, *i.e.*, the smaller the distance is, the larger the $Q$ is. If no other effects, the local moment $m$ shall equal to $Q$ or 2-$Q$. However, the hybridization also lowers the kinetic energy of the LUMO electrons, which competes with the onsite repulsion that drives spin splitting. If the interlayer interaction is too strong, the molecular frontier energy levels become metallic-like and the local moment tends to vanish. Different from the original Anderson model, here the local states are molecular frontier orbitals rather than transition metal *d* or *f* orbitals. The Coulomb repulsion U of the LUMO, mainly composed of *p* orbitals, is much smaller than that of *d* or *f* orbitals, and therefore the existence of local moment is very sensitive to the substrate.

For ML-PTCDA/Pb [blue curves in Fig. 1(c)], when *d* is less than 3.6 Å, *Q* exceeds one, and *μ* starts to decay. However, *m* decays much faster than the way *Q* increases and vanishes at *d* = 3.3 Å when *Q* is only 1.4e. The violation of the linear relationship between Q and $m_{\mu}$ is caused by the enhancement of the LUMO state broadening as PTCDA approaching the substrate. The partial density of states plots at different distances in Fig. 1(d) clearly shows the broadening effect.

The difference between the Pb and Ag substrates is the stronger hybridization between the PTCDA LUMO state and the Ag states, as compared to Pb. The partial densities of states of the PTCDA/Ag(111) system at four different *d* are plotted in Fig. S3(b). The occupied PTCDA LUMO states are significantly broader than the HOMO states at all distances, while on Pb(111) the broadening effects on the LUMO and HOMO are similar [Fig. 1(d)]. Such a strong hybridization makes the molecular states metallic-like, and thus no spin polarization occurs.

(3) **Examine the moiré modulation by the lateral shift of the PTCDA unit cell on Pb(111).** In order to qualitatively examine the modulation effect of moiré superstructure, we carried out calculations for a series of laterally shifted adsorption positions of the PTCDA unit cell. We found that the equilibrium interlayer separation, charge transfer, and magnetic moment can vary along with the changes of registries in the in-plane direction. Magnitudes of charge transfer and moment were obtained for each registry at its equilibrium interlayer distance. As shown in Fig. S3(c), there are two extreme cases with equilibrium *d* = 3.33 Å and 3.43 Å, respectively. For the case with *d* = 3.33 Å, we find a magnetic moment of *m* = 0.0 $\mu_B$ per molecule and a charge transfer of *Q* = 1.35e per molecule, while for the case with *d* = 3.43 Å, we find *m* = 0.34 $\mu_B$ and *Q* = 1.12e per molecule. Though these numbers are not expected to be highly accurate due to the approximate DFT and VDW forms we used, we expect that the calculated trend should be valid.

Though we believe the oscillation of Kondo temperature indeed results from the variations of the local interlayer registry (*i.e.*, adsorption site), just like other moiré induced super-periodicities. In our sample system, the effects of adsorption sites on Kondo shall be more complex than in other cases with the moments residing in *d*/*f* electrons. This is due to the fact that in our sample, not only the spin-exchange interaction but also the net magnetic moments are sensitive to the interlayer distance. For instance, in the above two configurations, though the exchange coupling may enhance when the *d* changes from 3.43 Å to 3.33 Å, the magnetic moment can completely diminish at *d* = 3.33 Å.

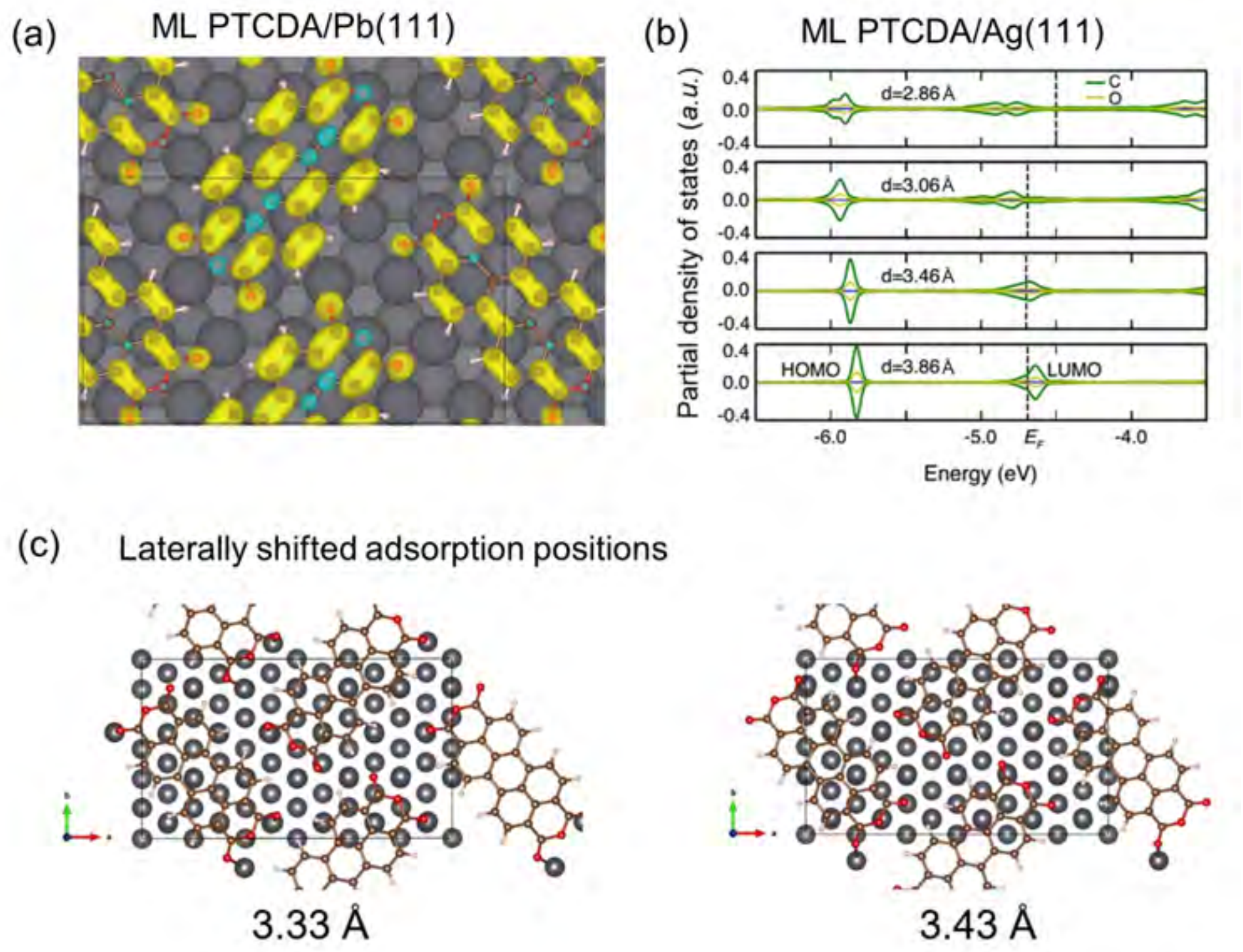

**Fig. S3** (a) Spin density of the PTCDA monolayer on Pb(111) surface at their equilibrium distance. Yellow: spin up; Blue: spin down. (b) PDOS of PTCDA layer on Ag(111) as a function of separation distance. (c) Schematic models of the two extreme cases where *d* = 3.33 Å and 3.43 Å, respectively, when a lateral shift is applied to the ML PTCDA/Pb system.

## 4. Fano fitting of Kondo resonance and the temperature dependence of HWHM (Γ)

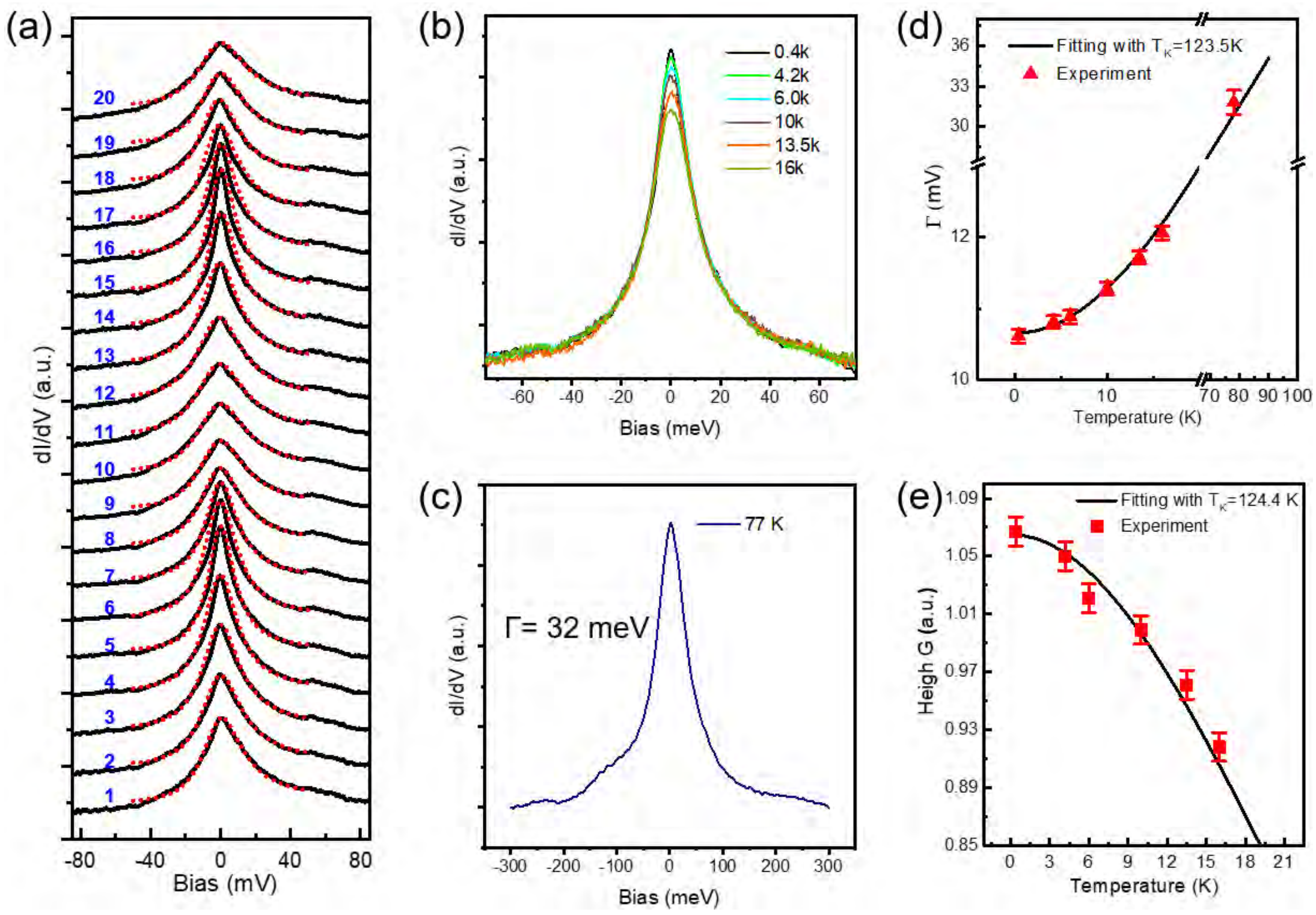

**Fig. S4** (a) Experimental spectra in Fig. 2(c) (balck) and the corresponding Fano fits (red). Molecule index numbers are labeled. (b) d*I*/d*V* spectra taken at various sample temperature (feedback opened at $V_B = 80$ mV, $I_t = 100$ pA and $V_{rms} = 1$ mV) on molecule #7. (c) d*I*/d*V* spectrum taken at $T_s = 77$ K (feedback opened at $V_s = 300$ mV, $I = 150$ pA and $V_{rms} = 3$ mV). (d) Γ as a function of the temperature. (e) Heights of the differential conductance peak at zero-bias $G_{V=0}$ versus temperature. The solid line is a fit to $G_{V=0(T)}$ by using empirical equation $G_{V=0} = G_0((T/T_K^*)^2+1)^{-s}$.

The zero-bias peaks observed in the d*I*/d*V* spectroscopy can be fitted with the Fano lineshape formula [25]:

$$\frac{dI}{dV} = H\frac{(q+\varepsilon)^2}{1+\varepsilon^2} + \rho_0, \text{ with } \varepsilon = \frac{eV - E_K}{\Gamma},$$

where $H$ is the magnitude, $\rho_0$ is the background signal, $E_K$ is the energetic position, $q$ is the Fano factor, and Γ is the half-width at half-maximum. The experimental spectra for all 20 molecules in Fig. 2(c) are present in Fig. S4(a), and the corresponding fitting curves are displayed as well. The Γ value for each molecule can be extracted from the fitting. For the present 20 molecules, we obtained Γ from 7.3 to 19.3 meV, which correspond to Kondo temperatures $T_K$ = 84 K~ 223 K.

In Fig. S4(b), we show the d$I$/d$V$ spectra taken on the mid-topographic height molecule #7 with the $T_S$ varying from 0.4 K to 16 K. One characteristic of the Kondo resonance is its temperature dependence, *i.e.*, the intrinsic width Γ grows linearly with temperature for higher temperatures whereas it saturates at a finite value at low temperature [26]. The function of Γ(T) can be expressed as $2\Gamma = \sqrt{(\alpha k_B T)^2 + (2K_B T_K)^2}$, where $T_K$ is the Kondo temperature ($\Gamma = T_K k_B$) and $\alpha$ represents the slope ranging from 5-10 in previous studies [26]. The equation above describes only the broadening of the Kondo feature itself and does not account for the instrumental broadening effects from the tip Fermi edge and the lock-in modulation voltage.

Following the approach in [27,28], we include the instrumental broadening effects by a simple quadratic approximation: $2\Gamma = \sqrt{(3.2k_B T_{tip})^2 + (\alpha k_B T)^2 + \left(2\sqrt{2}eV_{rms}\right)^2 + (2K_B T_K)^2}$, where $V_{rms}$ is the lock-in bias modulation and $T_{tip}$ is the tip temperature (set being equal to $T_S$ in our calculations). With $T_K$=123.5±4 K and $\alpha$ = 7.2, we obtain a good fitting to the experimental data points [Fig. S4(d)]. Moreover, the spectrum taken at $T_S$ = 77 K gives Γ = 32 meV [Fig. S4(c)], which agrees reasonably well with the present fitting. Note the measurement at 77K was taken on a molecule with the same index within the moiré periodicity (not on the same sample region).

In addition, the reduction of the zero-bias peak height ($G_{V=0}$) with increasing $T_S$ follows the empirical expression for a spin-1/2 system: $G_{V=0} = G_0((T/T_K^*)^2+1)^{-s}$, where the coefficient $s$ = 0.22 and $T_K^* = T_K/[\sqrt{2}\ (2^{1/s}-1)^{1/2}]$ [18,29]. Fig. S4(e) shows the fitting results for the data points extracted from Fig. S4(b), yielding the $T_K$ = 124.4±2 K, which agrees well with the value determined by the Γ(T) fitting. Note that different forms of the definition of $T_K$ may be found in the literature [30]. Here we have adopted the definition $T_K$ = HWHM/$k_B$. Therefore, a coefficient of $\sqrt{2}$ is used in the fitting of $G_{V=0}$(T).

Besides the characteristic temperature dependence shown above, other experimental facts corroborate the formations of magnetic moments as well: 1) The zero-bias peak (ZBP) only occurs in monolayer PTCDA, but vanishes in a single molecule. This rules out the possibility that the ZBP comes from the LUMO states of the molecule. 2) As shown in Fig. 3(d), the $T_C$ of Pb film was suppressed by ~ 20% by ML molecules (with a thickness only of ~ 0.15 nm). If it's a pure proximity effect from a non-magnetic metal (such as Ag), more than 8ML (~2 nm) of Ag(111) film is needed to produce the same suppression [31].

## 5. Additional data of twist angle dependent moiré modulations

**(1) Kondo and SC variations along a dark row:** Mapping of Kondo resonances along the dark row-*1* and selective spectra are displayed in Fig. S5(a) and S5(b), respectively. A much smaller variation in Γ is obtained with Γ= 8.6 meV, 9.7 meV , 9.9 meV for molecule #1, #5, and #7, respectively. Figure S5(c) shows spatial variations of $\Delta^*$ along the dark row-1. The minimum and maximum values of $\Delta^*$ are 0.70 meV and 0.91 meV, respectively. In Fig. S5(e), we plot the Γ and $\Delta^*$ for each molecule along the dark row-*1.* Though the magnitudes of Γ and $\Delta^*$ are still strictly correlated, the moiré oscillation is not conspicuous in the dark row.

In addition, oscillations of the line shape symmetries (*i.e.,* Fano factor *q*) are observed in our spectroscopic measurements as well. According to Fano's inference model, the magnitude of *q* is given by the ratio between the tunneling probabilities of two channels at the resonance energy, $t_1$ and $t_2$ (*i.e.,* $q \propto t_2/t_1$), where $t_1$ is the tunneling probability into metal continuum states, $t_2$ is the probability directly into the impurity state [26]. The sign of *q* represents the phase shift of the quantum interference between the two tunneling channels [32]. When q = +∞, the resonance will show as a Lorentzian peak; and |q| = -∞ represents a Lorentzian dip. Generally, the smaller the |q|, the more asymmetric the line shape. The sign of *q* presents the polarity of the asymmetry.

In Fig. S5(f) and S5(g), we plot the 1/*q* for each molecule along the bright row-*I* and the dark row-*1*, respectively. The corresponding height profiles (green dashed line) are displayed. A clear moiré correlated oscillation can be seen in the spatial dependence of 1/*q* from near zero to finite negative values (≥ -0.1) along the dark row. However, along the bright row, 1/*q* remains positive and oscillates between +0.01 and +0.18. In our case, |1/*q*| is much smaller than one. This means the tunneling into the impurity states dominates over the continuum channel (*i.e.,* into the Pb film). It is worth noting that in the bright row, the peak of 1/*q* occurs at the locations with medium *d* (*e.g.,* #7- #10), other than the locations with the lowest *d*. Combining with the measurements of bound states, we found that in our sample system, the Kondo and bound states asymmetries are not strictly correlated. This is different from the recent observation on FeTpyP-Cl/Pb [33].

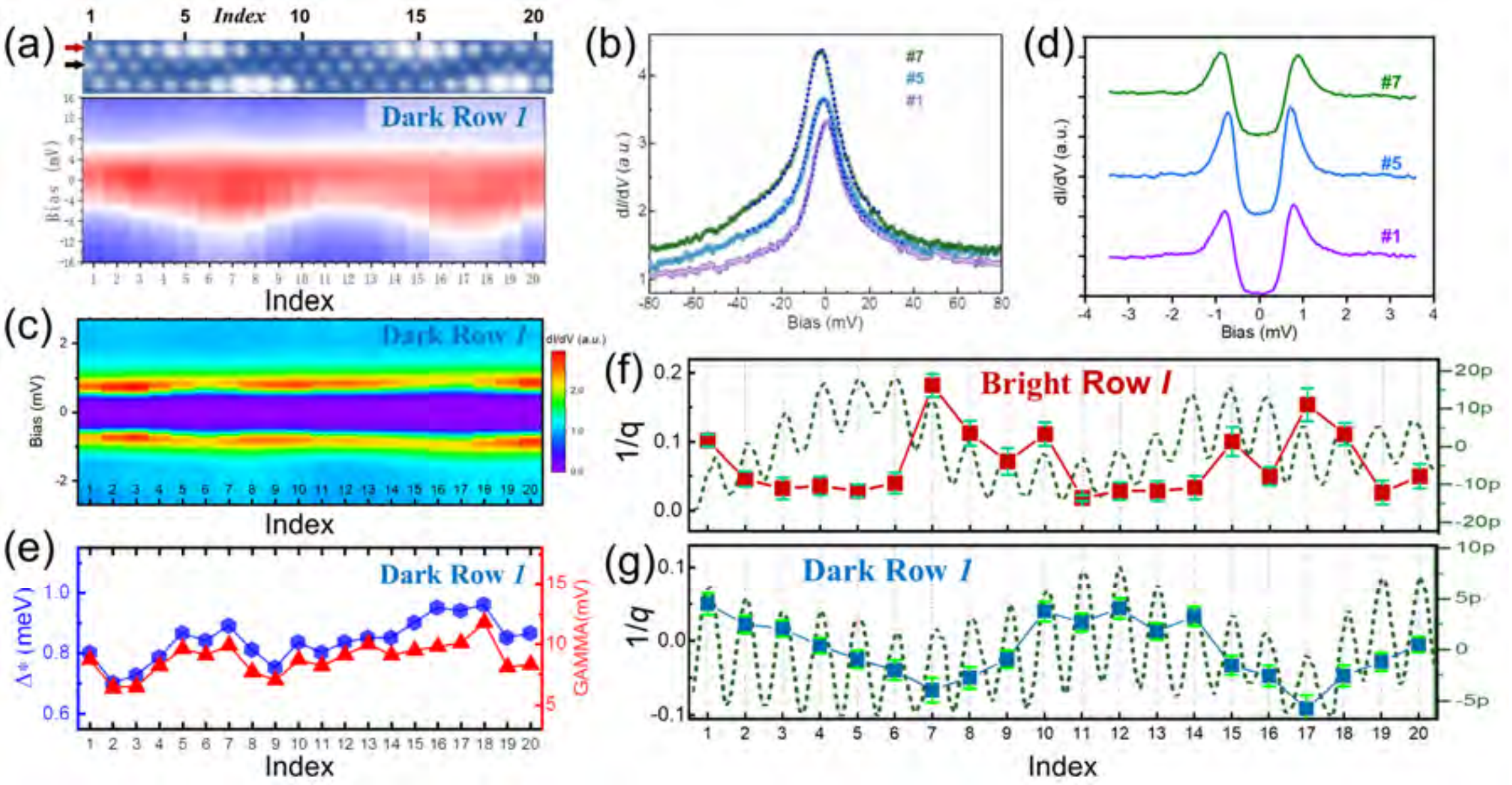

**Fig. S5** (a) False-color images of a series of d*I*/d*V* spectra taken along dark row-*1* (marked by a black arrow in the corresponding topographic image) with an energy range from +16mV to -16 mV. The red arrow indicates the bright row-I studied in the main text. (b) Representative spectra from (a) with their index numbers labeled. Spectra are slightly shifted for clarity. The dotted curves are the Fano fits. The fitting parameters are: Γ = 8.7 meV, $q$ = 20 for #1; Γ= 9.6 meV, $q$ = -40 for #5; Γ= 9.9 meV, $q$ = -15 for #7. All spectra were acquired at 0.4 K, and a magnetic field of 0.5 T was applied to suppress the superconductivity. (c) False-color image of the pairing gap measurements taken along the dark row-*1*. Feedback opened at $V_s$ = 3.5 mV, $I$ = 100 pA and $V_{rms}$ = 50 μV. (d) Selected spectra from (c) with their index number labeled as shown. A normal Pt/Ir tip was used here. (e) The variations of Γ and Δ* along the dark row-*1*. (f) and (g) The spatial variations of 1/$q$ along bright row-*I* and dark row-*1*, respectively. The corresponding height profiles (green dashed line) are displayed.

**(2) Moiré modulations of Kondo screening when $\theta = 13°$.** In Fig. S6(a), we show the modulation of Kondo resonance in the moiré superlattice with $\theta = 13°$. An oscillation correlated with the moiré periodicity is also observed. Similar to the situation of $\theta = 11°$, the peak value of Kondo temperature occurs at the location with the lowest topographic height, while the minimum value occurs at the highest points. Three representative spectra with the corresponding Fano fittings are displayed in Fig. S6(b). The maximum of Γ is found to be 14.9 meV, and the oscillation amplitude is about 6 meV [inset in Fig. S6(a)], leading to an averaged Γ of 10.4 meV. The corresponding averaged Γ in $\theta = 11°$ is 13.3 meV. Thus, we demonstrated that both the overall screening strength and its spatial oscillations in such a 2D impurity lattice can be effectively tuned by the interlayer twist-angle.

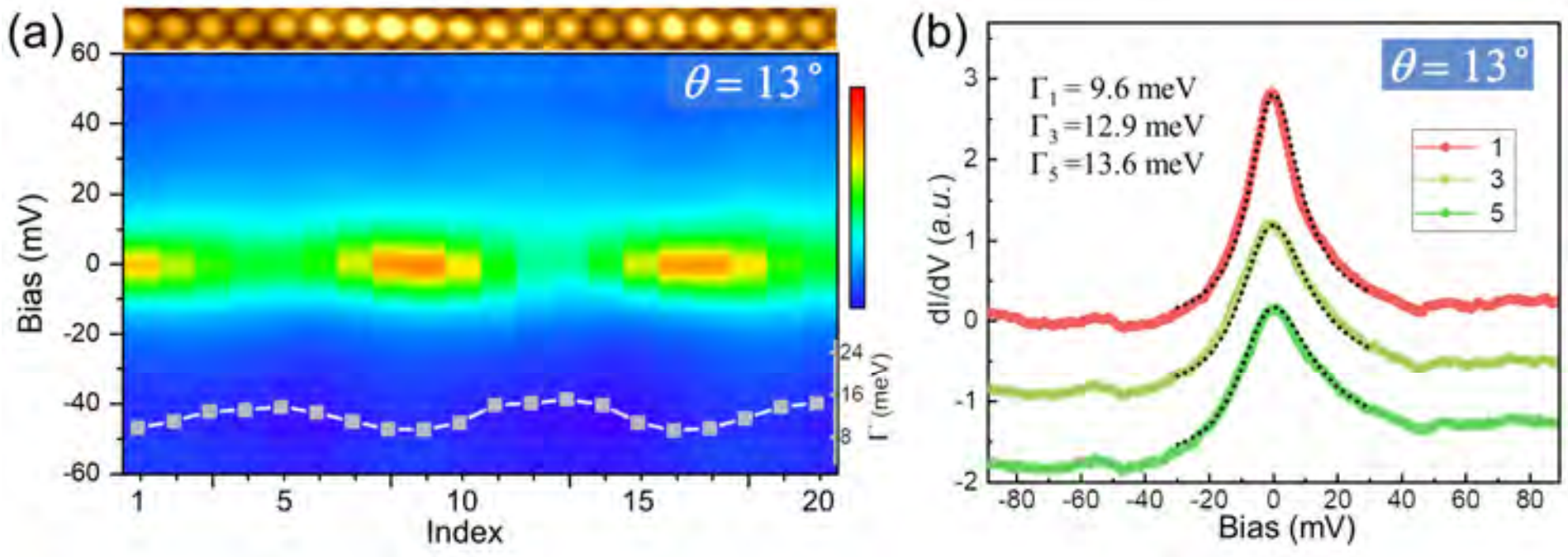

**Fig. S6** (a) False-color images of d$I$/d$V$ spectra taken along a bright row in the moiré superlattices with $\theta$= 13°. The insert shows the corresponding moiré modulations of Γ (grey). (b) Three representative spectra from (a) with the index numbers labeled as shown. The Γ values are marked. Slight vertical shifts are applied for clarity. The dotted black curves are the Fano fittings. All spectra were taken at 0.4 K with a small magnetic field applied to suppress the pairing gap. Feedback opened at $V_s$ = 95 mV, $I$ = 100 pA and $V_{rms}$ =2.5 mV.

## 6. Analyses of the tunneling spectroscopy of bound states

(1) **Deconvolution of the tunneling spectra.** Following the procedures as below, we achieved the numerical deconvolution of the d$I$/d$V$ spectra of PTCDA/Pb taken with a superconducting Nb tip. All experimental data analyzed in this section were taken at the sample temperature reading of 0.4 K.

In principle, the tunneling spectra is a convolution of the tip DOS and the sample DOS :

$$\frac{dI}{dV}(V) = \int_{-\infty}^{+\infty} \left\{ \left[\frac{\partial}{\partial V} n_{tip}(E+eV)\right] [f_{FD}(E) - f_{FD}(E+eV)] - [\frac{\partial}{\partial V} f_{FD}(E+eV)] n_{tip}(E+eV) \right\} n_{sample}(E) dE \qquad \text{(Eq. S1)}$$

where $n$ represents the DOSs, $f_{FD}$ is the Fermi-Dirac function.

We first performed the simulation for a spectrum taken with a normal tip on bare Pb film. The DOS function of the normal tip ($n_{tip}$) can be considered as a constant. The superconducting sample (or tip) has a BCS DOS as $n_{BCS} = n_F \left| Re\left(\frac{E+i\gamma}{\sqrt{(E+i\gamma)^2-\Delta^2}}\right)\right|$, where Δ is the order parameter, and $\gamma$ is the Dynes parameter representing the intrinsic gap broadening [34,35]. The least $\chi^2$ method is used to fit the convoluted d$I$/d$V$ function to the experimental spectrum, giving the fitting parameters as $\Delta_{Pb}$ = 1.35 meV, $\gamma_{Pb}$ = 6.2×10$^{-2}$ meV, and the effective electron temperature $T_{eff}$ = 0.63 K (comprised in the $f_{ED}$ function). The fitting result is plotted in Fig. S7(a).

With the acquired sample DOS and $T_{eff}$, we perform a similar fitting for the spectrum taken with an Nb tip on Pb film [Fig. S7(b)]. The BCS DOS function of the tip can be determined as $\Delta_{tip}$ = 1.47 meV, $\gamma_{tip}$ = 9.0×10$^{-2}$ meV. The small dyne parameters indicate the superconducting systems are rather pure. The finite value of $T_{eff}$ (0.63 K), which is close to the reading of sample temperature (0.4 K), manifests the state-of-the-art energy resolution of our STM.

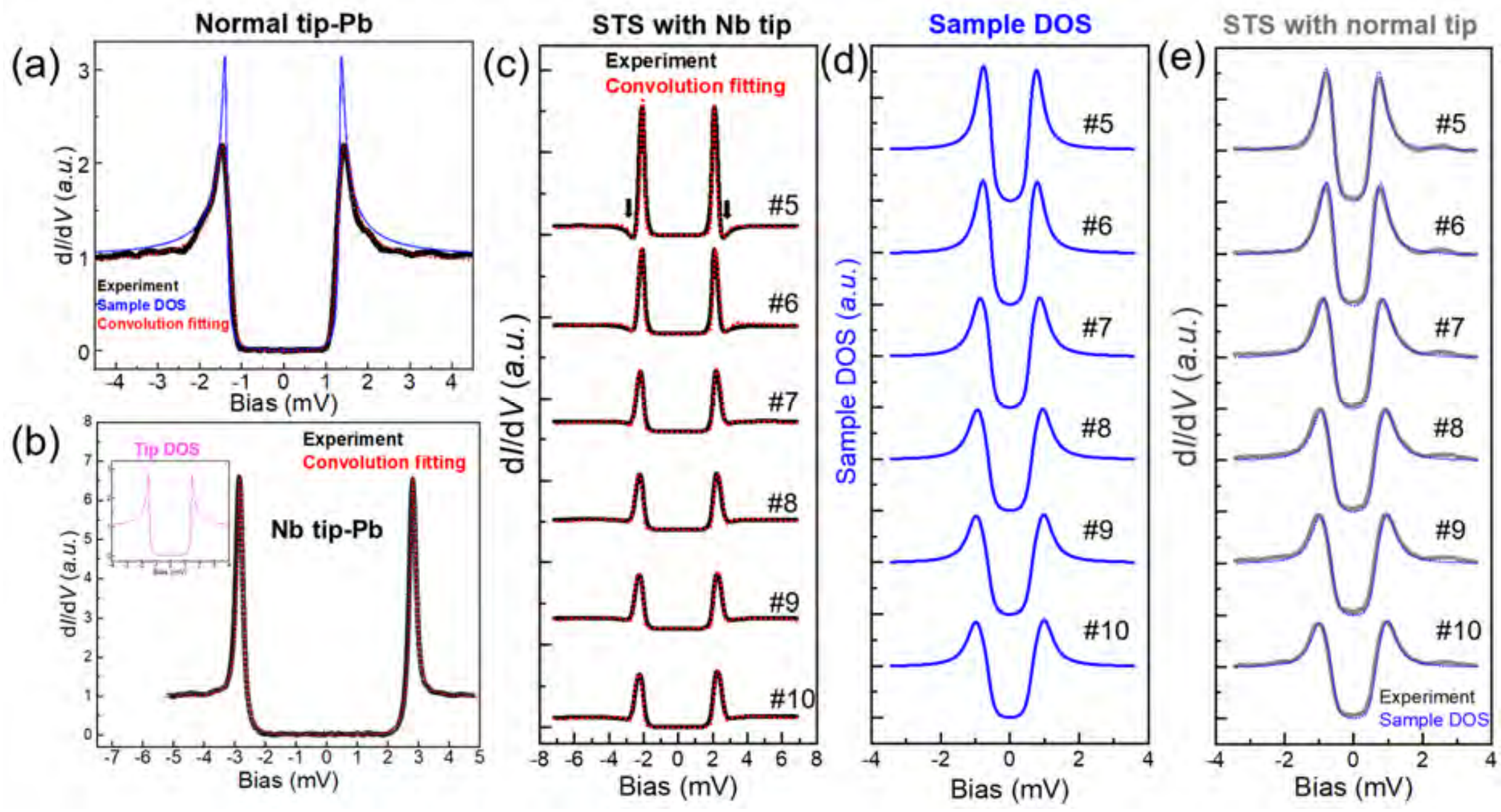

**Fig. S7** (a) Experimental spectrum of bare Pb taken with a normal tip (black). The blue curve represents the BCS DOS of the Pb film, and the red dotted curve is the convolution fitting with a constant tip DOS. (b) The experimental spectrum of bare Pb taken with a superconducting Nb tip (black). The red curve is the convolution fitting by using the sample DOS as shown in (a) and the tip DOS in the inset. (c) Experimental spectra with the same Nb tip for molecule #5-#10 (black), and the corresponding deconvoluted sample DOSs are displayed in (d). The convolution fittings [by using the sample DOS in (d) and the tip DOS in (b)] are present in (c) as red dotted lines, showing good consistencies with the experimental spectra. (e) shows experimental spectra taken with a normal tip at the same positions within a moiré periodicity (gray). The sample DOSs in (d) are overlaid for comparison. All experimental measurements here were taken at the sample temperature reading of 0.4 K.

The same Nb tip is used to measure the PTCDA/Pb sample. Therefore, the parameters $\Delta_{tip}$, $\gamma_{Pb}$, and $T_{eff}$ can be kept fixed in the subsequent deconvolution fitting. To simulate the DOS of PTCDA/Pb, we use the function as [36]:

$$n_{sample}(E) = \frac{1}{e^{\frac{\Delta-|E|}{\delta}}+1} + \frac{A_1}{1+\left(\frac{E-\Delta^*}{w}\right)^2} + \frac{A_2}{1+\left(\frac{E+\Delta^*}{w}\right)^2} \qquad \text{(Eq. S2)}$$

This model function consists of a symmetric gap (first item) and two Lorentz peaks that simulating the bound states. $A_1$ and $A_2$ represent the heights of the two peaks, *w* and $\Delta^*$ are the HWHM and the energy location of the peaks, respectively. The simulated sample DOSs for molecules from #5 to #10 are shown in Fig. S7(d). The extracted $\Delta^*$ are plotted in Fig. 2(c). The bandwidth of bound states (2*w)* varies from 0. 21 meV to 0.29 meV for #5-#10. The convoluted d*I*/d*V* functions are plotted as red dotted curves in Fig. S7(c), showing good agreements with the experimental curves (black). In Fig. S7(e), we present the spectra taken by a normal tip at the same positions within a moiré periodicity (not in the same region). They resemble the deconvoluted sample DOS well except for the instrumental broadening; thus, confirm the validity of our deconvolution process and the reproducibility of measurements.

(2) **Origin of the dip feature.** With a finite $\Gamma >> \Delta$, the energy levels of bound states are usually close to the superconducting gap edge; thus, it will be difficult to directly resolve them as in-gap peaks isolated from the gap edge even with a superconducting tip. Note that in some hybrid M-SC systems, there can be only a quenching of the BCS gap in STS measurements [37,38]. Following discussion helps us to exclude the latter possibility, hence confirms the existence of bound states in our system.

In the experimental spectra taken by a superconducting tip on the PTCDA/Pb, dips of the conductance are observed right above the excitation peaks, as shown in Fig. 3(a) and Fig. S7(c). The depth of the dip increases gradually from #10 to #5, and a negative differential conductance is obtained in spectrum #5.

In Fig. S8(a), areas of the excess spectral weight of the peak and the deficiency of the spectral weight in the gap region are marked $S_1$ and $S_2$, respectively. The theoretical sample DOSs are plotted in Fig. S8(b) based on Eq. S2, with $S_1/S_2$ varied from 2.85 to 0. For demonstration purposes, we use $A_1$=$A_2$ (*i.e.,* symmetric spectra weights) in the following discussion. The fitted tunneling spectra for various $S_1/S_2$ are present in Fig. S8(c). All other fitting parameters and the tip DOS are the same. It is clearly seen that the depth of the dip feature depends on the ratio of $S_1/S_2$. The tip DOS used here is a pure BCS one with its $S_1/S_2$ ~ 1. The dip feature results from the difference in the $S_1/S_2$ ratio between the sample and tip. The larger the difference is, the deeper the dip feature. The observations of the dip features confirmed the formations of bound states. Moreover, the consistency of the experimental data ($\Delta^*$ *vs.* $\Gamma$) with the Matsuura model [Fig. 3(d)] also supports the interpretation that the particle/hole excitations are the bound states.

We emphasize that the advantages of using the Nb tip in our studies are two-fold: i) the instrumental broadening can be qualitatively addressed and thus can be excluded from the reasons causing the large bandwidth of the bound states. ii) The dip features help us to identify the existence of bound states.

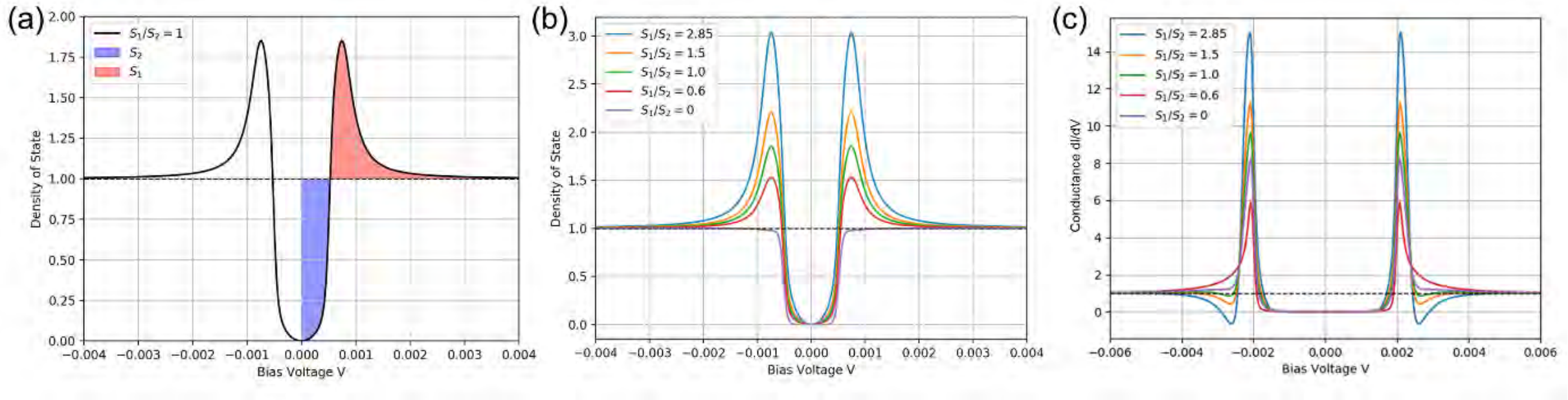

**Fig. S8** (a) Schematic to show the definitions of $S_1$ and $S_2$. (b) Series of simulated DOS based on the Eq. S2. The $S_1/S_2$ varies from 2.85 to 0, with other parameters fixed. (c) Simulations of the d$I$/d$V$ spectra for the sample DOS in (b). The tip DOS is the same one used in Fig. S7.

(3) **Calculation of $N_F|J|$ and $N_FK_U$.** Based on Schrieffer's theory [39], in the strong coupling regime, the $N_F|J|$ and $N_FK_U$ can be calculated by the following formulas.

$$\frac{\Delta^*}{\Delta_0} = \mathrm{sgn}(c_+c_-)\frac{c_+c_- - 1}{\sqrt{(c_+{}^2+1)(c_-{}^2+1)}}, \qquad \text{(Eq.S3)}$$

$$\frac{Z^+}{Z^-} = \frac{\overline{Z} - \delta Z}{\overline{Z} + \delta Z}, \qquad \text{(Eq.S4)}$$

where $c_\pm = c_j \mp c_u$, $c_j = (J/\pi N_F)/(J^2 - K_U{}^2)$, $c_u = (K_U/\pi N_F)/(J^2 - K_U{}^2)$,

$$\overline{Z} = (N_F\Delta_0)2\pi\left|c_j\right|[(c_j{}^2+c_u{}^2)+(c_j{}^2-c_u{}^2)^2]/[(c_+{}^2+1)(c_-{}^2+1)]^{3/2},$$

$$\delta Z = (N_F\Delta_0)\,\mathrm{sgn}(U)4\pi\left|c_u\right|c_w{}^2/[(c_+{}^2+1)(c_-{}^2+1)]^{3/2}.$$

Here, $\Delta^*$ is the energy of bound states, $\Delta_0$ is the order parameter of superconductivity, $Z^{(+)}$ and $Z^{(-)}$ are the intensities of electron- and hole- like excitations. We use $\Delta_0$ =1.06 meV from the Matsuura fitting. By substituting the experimental values of $\Delta^*$ and $\frac{Z^{(+)}}{Z^{(-)}}$ into Eq.S3 and Eq.S4, we can solve $N_F|J|$ and $N_FK_U$. In our analysis, we considered $N_F|J|$ as an integrated variant, and the same consideration is applied to $N_F K_U$. Note that from the experimental observation of $Z^{(+)} < Z^{(-)}$, we can determine that $K_U > 0$, while the sign of $J$ cannot be determined from the equations directly. Along the bright row-*I*, $N_F K_U$ is always a small value (from 0.006 to 0.057). We do not observe obvious oscillation of $K_U$ along with the moiré periodicity, likely due to its relatively small amplitude.

## 7. Delocalization of the magnetic moments: 2D nature of the bound states and Kondo

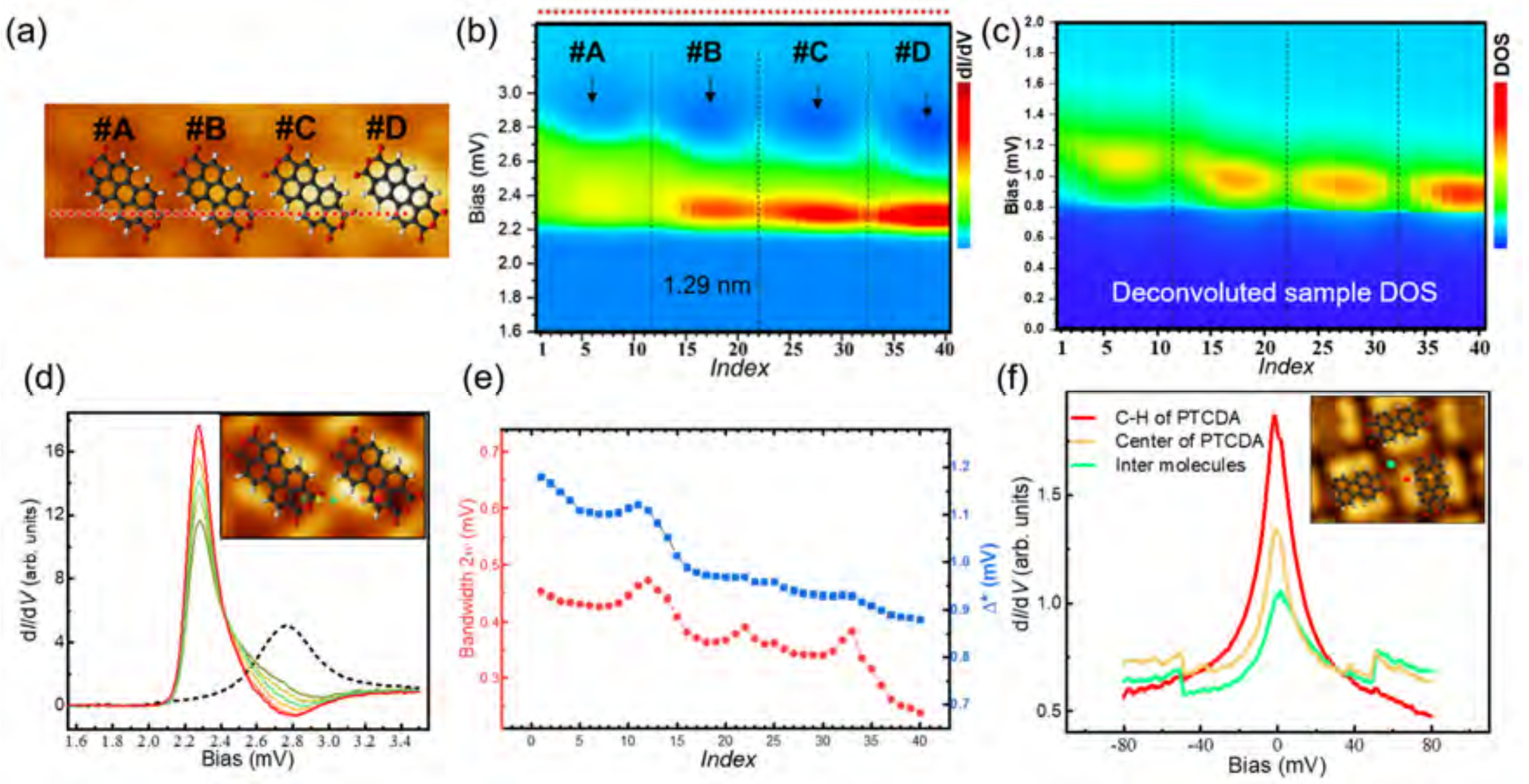

**Fig. S9** (a) A STM image with molecular models overlaid. Molecule #D has the highest morphology within a moiré periodicity. (b) False-color image of d*I*/d*V* spectra along the red path in (a). The spatial step size is 1.3 Å, 40 spectra in total. Only particle-excitation peaks are displayed. A superconducting Nb tip was used here. (c) False-color image of sample DOS obtained by numerical deconvolution. (d) Selected d*I*/d*V* spectra of the bound states across two nearby molecules. Locations are labeled in the inset with corresponding colors. The green curve is taken at the middle site between two molecules. A spectrum of bare Pb is shown for comparison (black curve). (e) The bandwidth 2*w* of bound states (red) and the $\Delta^*$ (blue) extracted from all the 40 spectra. (f) The measurements of Kondo resonance at three typical locations as labeled in the inset by the corresponding colors. The green one was taken on the middle site between two molecules exhibiting a pronounced Kondo feature are well. (b) $T_S$ = 0.4 K, $V_s$ = 3.5 mV $I$ = 100 pA, $V_{rms}$= 50μV; (f) $T_S$ = 4.2 K with $V_s$ = 80 mV $I$ = 200 pA, $V_{rms}$= 0.8 mV.

In Fig. S9(b), we show the d*I*/d*V* spectra mapping across four molecules in a bright row with a fine step size of ~1 Å (superconducting Nb tip was used) on PTCDA/22ML Pb. The corresponding topographic image is displayed in Fig. S9(a). Following the deconvolution procedure introduced in Section 6, we plot the sample DOS for all 40 spectra as a false-color image in Fig. S9(c). These results manifest clearly that the variation of bound states is *continuous* throughout the whole 2D layer. Note that the spectrum acquired at the location between two molecules [green spot in Fig. S9(d)] also exhibits the existence of the bound states confirming the delocalized nature of the bound states due to the molecular *sp* orbitals. In Fig. S9(e), we also plot the bandwidth 2*w* and Δ* for all the 40 locations, which are extracted from the convolution fittings (see Eq. S2). As discussed in the main text, when the excited states can overlap, an impurity band of the bound states can form with finite bandwidth. Indeed, we observed the magnitudes of 2*w* are abnormally large on PTCDA/22 ML Pb as well.

Moreover, Fig. S9(f) shows that the Kondo resonance exists even on the middle site between two nearby molecules, furthering confirming the delocalization of magnetic interactions. This is different from Ref. [36] where a superconducting gap of bare Pb recovers between two MnPc molecules due to "*a very narrow localization of the magnetic bound states*". In our system, however, the wavefunction of the magnetic moment is spatially extended, and the inter-molecular hybridization is non-negligible. The step features locating above ± 40mV shall correspond to the coupling of the electron orbitals with molecular vibrations, which were observed in other π-orbital Kondo systems (Ref. [18]).

## 8. Discussions about the QWS effect on the superconducting order parameters

(1) **Gap difference in the pristine 22 ML and 23 ML Pb.** It has been known that the quantum confinement effect only leads to a tiny change in superconducting $T_C$ between 22 ML and 23 ML bare Pb. In our measurements at 0.4K, it is indeed only a small difference in the superconducting gap. Figure S10(a) shows the d*I*/d*V* spectra acquired on 22ML and 23ML bare Pb with Nb tip. The peak-to-peak separations are 5.65 meV and 5.56 meV for 22 ML and 23 ML, respectively. Adopting the fitting results of $\Delta_{tip}$ = 1.47 meV (Fig. S7), we determine the pristine superconducting gap $\Delta_0^{22ML}$ =1.35 meV and $\Delta_0^{23ML}$ =1.31 meV, respectively. For ML-PTCDA/22ML-Pb hybrid bilayer, an ensemble of data sets of Γ *vs.* $\Delta^*$ [green dots in Fig. 3(d)], can also be fitted well to the theoretical model of Matsuura [40]. The asymptotic value of the superconducting order parameter $\Delta_0^{PTCDA/22ML}$ = is 1.23 meV at $\Gamma = \infty$, which is significantly large than $\Delta_0^{PTCDA/23ML}$ = 1.06 meV.

It was known that the 22ML Pb film has a relatively smaller work function compared to the 23ML one [41]. Therefore, a more significant charge transferring from Pb to the molecule is anticipated. As shown in our calculation, near the equilibrium adsorption distance, more net charges per molecule give rise to a decreasing in magnetic moments. In other words, the impurities concentration on 22ML Pb is less than the one on 23 ML Pb. It is indeed due to the unique origination of magnetic moments in our case (from interfacial charge transferring other than the intrinsic *d/f* unpaired orbital) that the quantum confinement effect in Pb film can play as a tuning knob for the magnetic impurities and the interplay between Kondo and SC.

(2) **Lateral proximity effect near the PTCDA ML edge**. In the fitting of $\Gamma$ *vs.* $\Delta^*$ by Matsuura's model, suppression of the superconducting order parameter is obtained. The fitting gives a $\Delta_0^{PTCDA/23ML}$ =1.06 meV, while the value measured on the bare 23 ML Pb film $\Delta_0^{23ML}$ is about 1.31 meV. Such suppression can be examined by the proximity behavior at the lateral interface between the molecule capped area and exposed Pb surface. Shown in Fig. S10(b) is an STM image showing a large-scale exposing Pb surface next to a PTCDA molecular layer. Using a Niobium tip, we acquired tunneling spectra across the lateral interface. As shown in Fig. S10(c), a reduction of the superconducting gap can be clearly seen even at location #5, which is ~ 65 nm away from the PTCDA island edge. At location #7-#9, the peak-to-peak separations are 5.56 meV, which is equal to $2(\Delta_{Nb}+\Delta_0^{23ML})$, indicating the pairing gap has been fully recovered. It is known that the typical length scales of the Kondo screening and bound states are ~ 2-5 nm [42]. The gradual reduction of the pairing gap observed here shall be due to the lateral proximity effect, which is analogous to our previous study on the Pb films with different thicknesses [43,44].

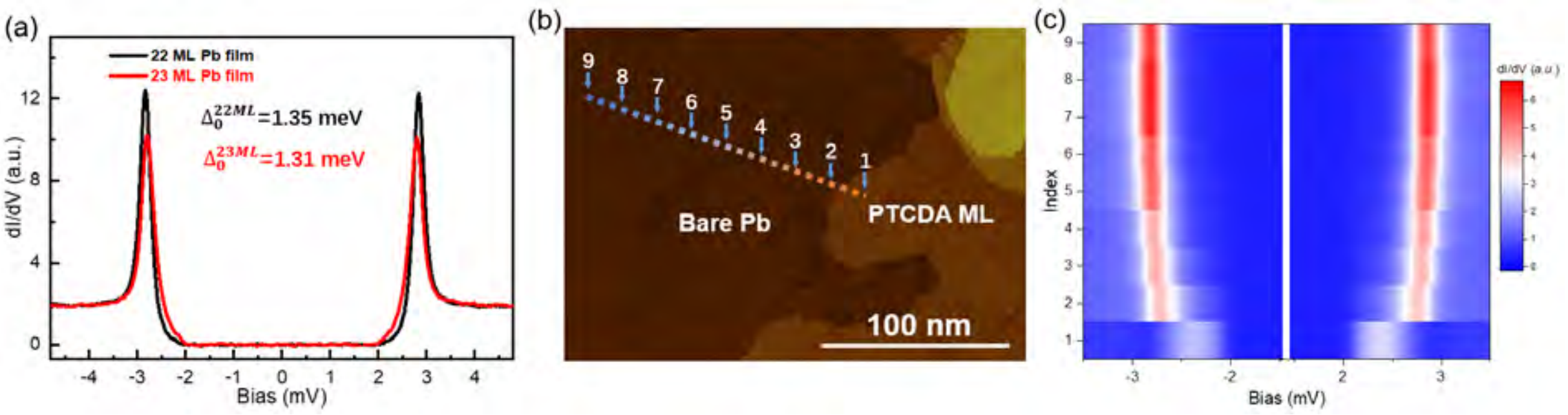

**Fig. S10** (a) d$I$/d$V$ spectra of bare 22ML and 23 ML Pb films measured with a superconducting Nb tip. (b) An STM image showing Pb surface partially covered with a PTCDA monolayer. $V_s$ = 0.1 V, $I$ = 10 pA, size 280×200 nm$^2$. (c) The spatial dependence of pairing gap. The spectra were taken at 0.4 K with an Nb tip along the dashed line in (b). The step size is 20 nm. Stabilization parameters $V_s$ = 5.0 mV, $I$ = 200 pA and $V_{rms}$ = 50 μV.